\documentclass[aps,pra,twocolumn,superscriptaddress]{revtex4}
\usepackage{amsfonts,amssymb,graphicx}
\usepackage{amsmath}
\usepackage[usenames,dvipsnames]{xcolor}
\usepackage[normalem]{ulem}
\usepackage{color}

\begin{document}

\title{Few-magnon excitations in a frustrated spin-$S$ ferromagnetic chain with single-ion anisotropy}
\author{Jiawei Li\footnote{These two authors equally contributed to the work.}}
\affiliation{Center for Quantum Technology Research, School of Physics, Beijing Institute of Technology, Beijing 100081, China}
\author{Ye Cao$^*$}
\affiliation{Key Laboratory of Advanced Optoelectronic Quantum Architecture and Measurements (MOE), School of Physics, Beijing Institute of Technology, Beijing 100081, China}
\author{Ning Wu}
\email{wunwyz@gmail.com}
\affiliation{Center for Quantum Technology Research, School of Physics, Beijing Institute of Technology, Beijing 100081, China}
\affiliation{Key Laboratory of Advanced Optoelectronic Quantum Architecture and Measurements (MOE), School of Physics, Beijing Institute of Technology, Beijing 100081, China}
\begin{abstract}
We study few-magnon excitations in a finite-size spin-$S$ chain with ferromagnetic nearest-neighbor (NN) interaction $J>0$ and antiferromagnetic next-nearest-neighbor (NNN) interaction $J'<0$, in the presence of the single-ion (SI) anisotropy $D$. We first reveal the condition for the emergence of zero-excitation-energy states. In the isotropic case with $\Delta=\Delta'=1$ ($\Delta$ and $\Delta'$ are the corresponding anisotropy parameters), a threshold of $J/|J'|$ above which the ground state is ferromagnetic is determined by exact diagonalization for short chains up to $12$ sites. Using a set of exact two-magnon Bloch states, we then map the two-magnon problem to a single-particle one on an effective open chain with both NN and NNN hoppings. The whole two-magnon excitation spectrum is calculated for large systems and the commensurate-incommensurate transition in the lowest-lying mode is found to exhibit different behaviors between $S=1/2$ and higher spins due to the interplay of the SI anisotropy and the NNN interaction. For the commensurate momentum $k=-\pi$, the effective lattice is decoupled into two NN open chains that can be exactly solved via a plane-wave ansatz. Based on this, we analytically identify in the $\Delta'-D/|J'|$ plane the regions supporting the SI or NNN exchange two-magnon bound states near the edge of the band. In particular, we prove that there always exists a lower-lying NN exchange two-magnon bound state near the band edge for arbitrary $S\geq 1/2$. Finally, we numerically calculate the $n$-magnon spectra for $S=1/2$ with $n\leq5$ by using a spin-operator matrix element method. The corresponding $n$-magnon commensurate instability regions are determined for finite chains and consistent results with prior literature are observed.
\end{abstract}

\maketitle

\section{Introduction}
\par Frustrated quantum spin systems with competing interactions can exhibit rich interesting phenomena due to the simultaneous existence of frustration and quantum fluctuations. In the past few decades, the spin-1/2 Heisenberg chain with ferromagnetic NN and antiferromagnetic NNN interactions has attracted considerable attention and has been thoroughly studied by using a variety of methods~\cite{Bader1979,Tonegawa1990,Chubukov1991,Krivnov1996,Dmitriev2006,Vekua2006,Vekua2007,Kuzian2007,Furusaki2007,Furusaki2008,Sudan2009}. The model is relevant to various quasi-one-dimensional magnetic materials such as $\mathrm{Rb_2Cu_2Mo_3O_{12}}$~\cite{Exp1} and $\mathrm{LiCuVO_4}$~\cite{Exp2}.
\par Theoretically, the NN-NNN spin chain (or the $J-J'$ chain in our notation) is simple enough and serves as a prototype for exploring novel quantum phases in more general frustrated magnetic systems. Besides its ground-state properties~\cite{Tonegawa1990,Krivnov1996,Dmitriev2006,Furusaki2008,Sudan2009}, of special interest is few-magnon excitations upon the fully polarized state~\cite{Chubukov1991,Kuzian2007,Furusaki2007,Furusaki2008}. In an early work, Chubukov studied the one- and two-magnon instability of a spin-1/2 $J-J'$ chain by using the bosonization technique based on the Dyson-Maleev transformation~\cite{Chubukov1991}. Kuzian and Drechsler mapped the two-magnon problem onto an effective tight-binding one and obtained the exact two-magnon excitation spectrum for infinite chains~\cite{Kuzian2007}. Kecke, Momoi, and Furusaki constructed a set of $n$-magnon Bloch states and calculated the $n$-magnon excitation spectra for $n\leq 4$ in a restricted Hilbert space~\cite{Furusaki2007}. The same method was subsequently used to calculate excitations up to $n=7$ and to identify the multimagnon bound states~\cite{Furusaki2008}.
\par Recently, there has been a resurgence of theoretical interest in few-excitations and their dynamics in quantum chains~\cite{Essler2012,Andrei,Balents,PRB2022,HJC2022,Pan2019,Guan2021}. This is mainly triggered by recent experimental advances in simulating spin-1/2~\cite{Fukuhara2013,Ketterle2020,Roos2023} and higher-spin~\cite{Ketterle2021,Ketterle2022} quantum magnetic models in cold-atom systems. We note that a uniaxial SI anisotropy term in spin-1 models was realized with ultracold atoms~\cite{Ketterle2021} and a long-ranged anisotropic Heisenberg model was recently realized using Floquet engineering~\cite{Roos2023}. Despite these experimental advances, multimagnon bound states in higher-spin $J-J'$ chains with SI anisotropy have been scarcely studied theoretically.
\par In this work, motivated by the above-mentioned experimental developments, we study theoretically few-magnon excitations upon the ferromagnetic state in a spin-$S$ periodic $J-J'$ chain with arbitrary $S$ and in the presence of the SI anisotropy. We first reveal the condition for the existence of zero-excitation-energy states and relate it to a threshold of $J/|J'|$ above which the ground state is ferromagnetic. By performing exact diagonalizations of short chains with $N\leq 12$ sites, we find that this threshold is always $J/|J'|=4$ for $S=1/2$, but shows size-dependence for $S>1/2$. Using a set of recently proposed exact two-magnon Bloch states for the finite-size XXZ chain~\cite{PRB2022}, we then map the two-magnon problem onto a single-particle one defined on an inhomogeneous open chain with both NN and NNN hoppings. Numerical solutions of the single-particle problem recover all the prior results for $S=1/2$~\cite{Chubukov1991,Kuzian2007,Furusaki2007,Furusaki2008}, including the identification of the two-magnon commensurate-incommensurate transition point in the lowest-lying excited state, the appearance of two-magnon bound states below the scattering continuum, etc. For $S>1/2$, the evolution of the lowest two-magnon excitation energy and the associated wave number with varying $J/|J'|$ behaves differently from the case of $S=1/2$. The SI anisotropy is found to have a large impact on the low-energy excitations.
\par To understand the emergence of bound states near the band edge, we note that for the commensurate momentum $k=-\pi$ the effective lattice is divided into two independent NN open chains. We solve the eigenvalue problem for these two decoupled NN chains by employing a plane-wave ansatz, from which we identify analytically the parameter regions supporting the two types of two-magnon bound states, i.e., the NNN exchange and SI two-magnon bound states (see Sec.~\ref{SecIV}). In particular, we rigorously prove that there always exists a lower-lying NN exchange two-magnon bound state in the $k=-\pi$ sector for arbitrary $S\geq 1/2$, regardless of all the parameters. These analytical results are expected to faithfully describe the spectrum structure near the band edge.
\par We also study $n$-magnon ($n\geq 3$) excitations in a spin-1/2 chain. Using a basis in which the NN XX interaction is diagonal, we present numerically exact calculations of the excitation spectra up to $n= 5$ in finite-size chains. The saturated magnetic fields and the associated number of magnons in the lowest excitation state are consistent with those obtained in a restricted Hilbert space~\cite{Furusaki2008}.
\par The rest of the paper is organized as follows. In Sec.~\ref{SecII}, we introduce the spin-$S$ $J-J'$ model and study the simplest subspace with only one magnon. We then introduce the exact two-magnon Bloch states and the plane-wave ansatz that will be used later. In Sec.~\ref{SecZEES}, we discuss the emergence of zero-excitation-energy states under certain conditions. In Sec.~\ref{SecIV}, we present detailed results about the two-magnon excitation and solve the problem for mode $k=-\pi$ semianalytically, with which we determine the emergence of two-magnon bound states near the band edge. In Sec.~\ref{SecV}, we focus on $n$-magnon excitations in the case of $S=1/2$. The exact excitation spectra for $n\leq 5$ are numerically calculated for finite chains. Conclusions are drawn in Sec.~\ref{SecVI}.
\section{Model and methodology}\label{SecII}
\subsection{Hamiltonian}
\par We consider a spin-$S$ homogeneous Heisenberg chain with both NN and NNN interactions
\begin{eqnarray}\label{Haml}
H&=&H_{\mathrm{NN}}+H_{\mathrm{NNN}}+H_{\mathrm{D}}+H_{\mathrm{B}},\nonumber\\
H_{\mathrm{NN}}&=&- J\sum^{N}_{j=1} (S^x_{j}S^x_{j+1}+S^y_{j}S^y_{j+1}+\Delta S^z_{j}S^z_{j+1})\nonumber\\
H_{\mathrm{NNN}}&=&- J' \sum^{N}_{j=1} (S^x_{j}S^x_{j+2}+S^y_{j}S^y_{j+2}+\Delta' S^z_{j}S^z_{j+2})\nonumber\\
H_{\mathrm{D}}&=&-D\sum^{N}_{j=1}  (S^z_{j})^2,~H_{\mathrm{B}}=-B\sum^{N}_{j=1}  S^z_{j},
\end{eqnarray}
where $\vec{S}_j=(S^x_j,S^y_j,S^z_j)$ are spin operators on site $j$ with quantum number $S\geq1/2$, $J$ and $J'$ measure the exchange interactions between NN and NNN spin pairs respectively with $\Delta, \Delta'>0$ the interaction anisotropies, $D\geq 0$ is the single-ion anisotropy strength and $B$ is an external magnetic field. Note that for $S=1/2$ the single-ion anisotropy term contributes only a constant $H_{\mathrm{D}}=-ND/4$. We therefore simply set $D=0$ in all the following discussions concerning $S=1/2$. For simplicity, we impose periodic boundary conditions $\vec{S}_j=\vec{S}_{N+j}$ and assume that $N$ is even and divisible by 4 (other cases can be similarly analyzed). The spin chain is translationally invariant under shifts by one lattice spacing.
\par  It is easy to see that the total magnetization $M=\sum_jS^z_j$ is conserved. We consider the case of $J>0$ and $J'<0$, where the antiferromagnetic NNN interaction induces a frustration~\cite{Chubukov1991,Furusaki2007}. We take the fully polarized state $|F\rangle=|S,S,\cdots,S\rangle$ as a reference state, which possesses an eigenenergy
\begin{eqnarray}\label{EF}
E_F&=&-NS^2(J\Delta +J'\Delta' +D)-NSB.
\end{eqnarray}
\par  The $n$-magnon subspace is spanned by all the spin configurations having $n$ spin deviations (with $S^-_j=S^x_j-iS^y_j$),
\begin{eqnarray}\label{j1j2jn}
|j_1,j_2,\cdots, j_n\rangle=\mathcal{C} S^-_{j_1}S^-_{j_2}\cdots S^-_{j_n}|F\rangle,\nonumber
\end{eqnarray}
where $\mathcal{C}$ is a suitable normalization constant and $1\leq j_1\leq j_2\leq\cdots \leq j_n\leq N$. The lattice translation operator $T$ is defined by the relation
\begin{eqnarray}
T|j_1,j_2,\cdots, j_n\rangle=|j_1+1,j_2+1,\cdots, j_n+1\rangle.\nonumber
\end{eqnarray}
It is obvious that $T^N=1$.
\subsection{One-magnon sector}
\par As a warm up, let us first study the single-magnon subspace. The $N$ one-magnon states are given by~\cite{PRB2022}
\begin{eqnarray}\label{1magnonBloch}
|\xi(k)\rangle=\frac{1}{\sqrt{N}}\sum^{N-1}_{n=0}e^{ikn}T^n|1\rangle,~k\in K_0
\end{eqnarray}
where the wave numbers $k$'s take values from the set
\begin{eqnarray}\label{K0}
K_0=\left\{-\pi,-\pi+\frac{2\pi}{N},\cdots,0,\cdots,\pi-\frac{2\pi}{N}\right\}
\end{eqnarray}
to guarantee the translational invariance of $|\xi(k)\rangle$, i.e., $T|\xi(k)\rangle=e^{-ik}|\xi(k)\rangle$. It is easy to check that $|\xi(k)\rangle$ is an eigenstate of $H$ with eigenenergy $E_F+\mathcal{E}_1(k)$, where
\begin{eqnarray}\label{E1magnonBloch}
\mathcal{E}_1(k)&=&-2S(2J'\cos^2k+J\cos k)+D(2S-1)+B\nonumber\\
&&+2S[J\Delta+J'(\Delta'+1)].
\end{eqnarray}
\par To study the instability of the ferromagnetic state $|F\rangle$, we define the spin gap $G_1$ as the energy difference between the lowest one-magnon state and $E_F$ in the absence of the magnetic field~\cite{Ueda2020}:
\begin{eqnarray}
G_1=\max\{ \mathcal{E}_1(k^{(\min)}_1)|_{B=0},0\},
\end{eqnarray}
where $k^{(\min)}_1$ is the wave number at which $\mathcal{E}_1(k)|_{B=0}$ reaches its minimum.
\par Since $\mathcal{E}_1(k)|_{B=0}$ is a quadratic function of $\cos k$, the wave number $k^{(\min)}_1$ is independent of the quantum number $S$ and the anisotropy parameters $\Delta$ and $\Delta'$ but depends only on the ratio $\mathcal{R}\equiv J/4|J'|>0$. As a result, $G_1$ exhibit different behaviors depending on whether $\mathcal{R}\geq 1$ or $\mathcal{R}< 1$.
\par i) $\mathcal{R}\geq 1$.
\par In this case we have $k^{(\min)}_1=0$ and
\begin{eqnarray}
\mathcal{E}_1(0)|_{B=0}&=& 2S[J(\Delta-1)+J'(\Delta'-1)]+D(2S-1).\nonumber\\
\end{eqnarray}
The one-magnon spin gap $G_1$ exactly vanishes for a spin-1/2 isotropic $J-J'$ chain with $\Delta=\Delta'=1$~\cite{Chubukov1991,Furusaki2007}, where $|F\rangle$ is degenerate with the lowest one-magnon state. However, this degeneracy is removed for higher spins if the SI anisotropy is present. In any case, $G_1$ is positive for $\Delta>1$ and $\Delta'<1$.
\par ii) $0<\mathcal{R}<1$.
\par In this case $k^{(\min)}_1$ takes the value such that $|\cos k^{(\min)}_1-\mathcal{R}|$ is the smallest. For finite $N$, $k^{(\min)}_1=0$ if and only if
\begin{eqnarray}\label{Rth}
\mathcal{R}> \cos^2\frac{\pi}{N}.
\end{eqnarray}
For large enough $N$, we have $k^{(\min)}_1\approx \arccos\mathcal{R}$, giving
\begin{eqnarray}
\mathcal{E}_1(k^{(\min)}_1)|_{B=0}&\approx&2SJ'[2(\mathcal{R}-\Delta)^2+(\Delta'+1-2\Delta^2)]\nonumber\\
&& +D(2S-1).
\end{eqnarray}
The condition for $G_1>0$ is
\begin{eqnarray}\label{Cd1}
\left(1-\frac{1}{2S}\right)D/|J'|> 2\mathcal{R}^2-4\mathcal{R}\Delta +(\Delta'+1).
\end{eqnarray}
For $S=1/2$ and $\Delta=\Delta'=1$, the above inequality can never be satisfied. Thus, it is necessary to introduce an easy-axis anisotropy or a nonzero magnetic field in order to search for a region of ferromagnetic phase for $0<\mathcal{R}<1$~\cite{Chubukov1991}. For $S>1/2$, Eq.~(\ref{Cd1}) can be fulfilled by choosing sufficiently large $D/|J'|$.  The required saturation field for $\Delta=\Delta'=1$ and $D=0$ is obviously
\begin{eqnarray}\label{Bsat}
B_{\mathrm{sat}}=S(J+4J')^2/4|J'|,
\end{eqnarray}
which is proportional to the quantum number $S$. A finite positive SI anisotropy $D$ can help lower the saturation field.
\subsection{Two-magnon Bloch Hamiltonians}
\par The two-magnon excitations of the $J-J'$ chain in the case of $S=1/2$ have been well studied by using various methods~\cite{Chubukov1991,Kuzian2007,Furusaki2007,Furusaki2008}. Here, we employ a set of recently proposed exact two-magnon Bloch states to investigate the two-magnon excitations for general $S$.
\begin{figure}
\includegraphics[width=0.51\textwidth]{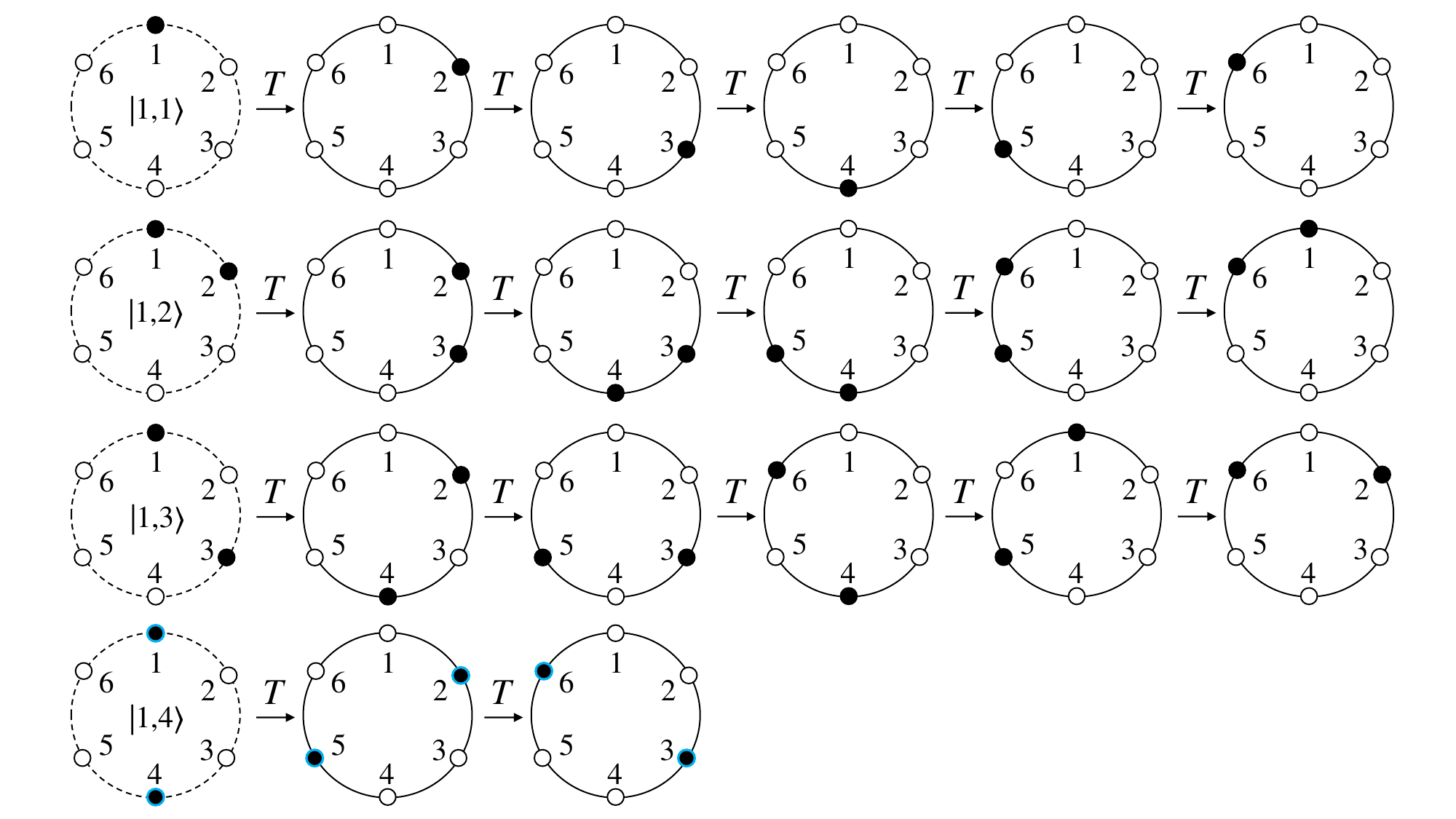}
\caption{For $N=6$ and $S>1/2$, the three parent states $|1,1\rangle$, $|1,2\rangle$, and $|1,3\rangle$ each generates five new states under the action of the translation operator $T$. However, the last parent state $|1,4\rangle$ generates only two new states. }
\label{Fig0}
\end{figure}
\par For $S>1/2$, the dimension of the two-magnon subspace is $\binom{N}{2}+N=N(N+1)/2$. There are two types of normalized two-magnon basis states in the real space,
\begin{eqnarray}
|i,j\rangle&=&\frac{1}{2S}S^-_iS^-_j|F\rangle,~1\leq i<j\leq N,\nonumber\\
|i,i\rangle&=&\frac{1}{2\sqrt{S(2S-1)}}(S^-_i)^2|F\rangle,~1\leq i \leq N.
\end{eqnarray}
Note that $|i,i\rangle$ is not defined for $S=1/2$. These states can be generated by successively applying the translation operator $T$ to the following parent states
\begin{eqnarray}
|1,1\rangle,|1,2\rangle,\cdots,|1,N/2\rangle,~\mathrm{and}~|1,N/2+1\rangle.\nonumber
\end{eqnarray}
Each of the first $N/2$ parent states generates $N-1$ additional states under the action of $T$, while the last one, $|1,N/2+1\rangle$, generators only $N/2-1$ additional states, see Fig.~\ref{Fig0} for an example with $N=6$.
\par We can linearly combine each parent state with its translated states to form a Bloch state labeled by the separation $r$ of the two spin deviations. Explicitly, for $r=0,1,\cdots,N/2-1$ we define~\cite{PRB2022}
\begin{eqnarray}\label{Bloch2m1}
|\xi_r(k)\rangle&=&\frac{e^{i\frac{rk}{2}}}{\sqrt{N}}\sum^{N-1}_{n=0}e^{ikn}T^{n}|1,1+r\rangle,
\end{eqnarray}
where $k\in K_0$. For $r=N/2$, we construct
\begin{eqnarray}\label{Bloch2m2}
|\xi_{\frac{N}{2}}(k)\rangle&=&e^{i\frac{Nk}{4}}\sqrt{\frac{2}{N}}\sum^{\frac{N}{2}-1}_{n=0}e^{ikn}T^{n}|1,1+\frac{N}{2}\rangle,
\end{eqnarray}
where $k\in K_1$ with (for even $N/2$)~\cite{PRB2022}
\begin{eqnarray}
K_1=\left\{-\pi,-\pi+\frac{4\pi}{N},\cdots,0,\cdots,\pi-\frac{4\pi}{N}\right\}.
\end{eqnarray}
The such constructed Bloch states are all normalized and translationally invariant, i.e., $T|\xi_r(k)\rangle=e^{-ik}|\xi_r(k)\rangle,~r=0,1,2\cdots,N/2$. We would like to mention that these set of Bloch states have been proposed for $S=1/2$ in the discussion of excitonic bound states in molecular chains~\cite{Ezaki1993}. For later use the complement of $K_1$ will be denoted as $K'_1$ such that $K_0=K_1\bigcup K'_1$.
\par For each $k\in K_1$, it can be shown by straightforward calculation that the $N/2+1$ ordered Bloch states $\{|\xi_0(k),\cdots,\xi_{N/2}(k)\}$ form a close set under the action of each individual term in the Hamiltonian. The two terms $H_{\mathrm{D}}$ and $H_{\mathrm{B}}$, as well as the Ising-coupling parts of $H_{\mathrm{NN}}$ and $H_{\mathrm{NNN}}$, do not involve spin flips and are all diagonal in the above basis. The XX-coupling part of $H_{\mathrm{NN}}$ was obtained in Ref.~\cite{PRB2022} as a tridiagonal matrix. For completeness, here we sketch how to evaluate the matrix elements of the XX-coupling part of $H_{\mathrm{NNN}}$. Let $\tilde{H}_{\mathrm{XX,NNN}}=\sum^N_{j=1}(S^+_jS^-_{j+2}+S^-_jS^+_{j+2})/2$, we choose $|\xi_{N/2-2}(k)\rangle$ as a representative Bloch state to calculate $\tilde{H}_{\mathrm{XX,NNN}}|\xi_{N/2-2}(k)\rangle$. We first calculate the action of $\tilde{H}_{\mathrm{XX,NNN}}$ on the parent state $|1,N/2-1\rangle$:
\begin{eqnarray}
&&\tilde{H}_{\mathrm{XX,NNN}}|1,\frac{N}{2}-1\rangle\nonumber\\
&=&\frac{1}{2}(S^-_{N-1}S^+_1+S^+_1S^-_3+S^-_{\frac{N}{2}-3}S^+_{\frac{N}{2}-1}+S^+_{\frac{N}{2}-1}S^-_{\frac{N}{2}+1})\nonumber\\
&&|1,\frac{N}{2}-1\rangle\nonumber\\
&=&S |\frac{N}{2}-1,N-1\rangle+S|3,\frac{N}{2}-1\rangle\nonumber\\
&&+S|1,\frac{N}{2}-3\rangle+S|1,\frac{N}{2}+1\rangle\nonumber\\
&=&S(1+T^2)|1,\frac{N}{2}-3\rangle+S(1+T^{-2})|1,\frac{N}{2}+1\rangle.\nonumber
\end{eqnarray}
By noting that $[\tilde{H}_{\mathrm{XX,NNN}},T]=0$ and $T^{N/2}|1,\frac{N}{2}+1\rangle=|1,\frac{N}{2}+1\rangle$, we have
\begin{eqnarray}
&&\tilde{H}_{\mathrm{XX,NNN}}|\xi_{N/2-2}(k)\rangle\nonumber\\
&=&\frac{e^{i(\frac{N}{2}-2)\frac{k}{2}}}{\sqrt{N}}\sum^{N-1}_{n=0}e^{ikn}T^{n}S(1+T^2)|1,\frac{N}{2}-3\rangle\nonumber\\
&&+\frac{e^{i(\frac{N}{2}-2)\frac{k}{2}}}{\sqrt{N}}\sum^{\frac{N}{2}-1}_{n=0}e^{ikn}T^{n}S(1+T^{-2})|1,\frac{N}{2}+1\rangle\nonumber\\
&&+\frac{e^{i(\frac{N}{2}-2)\frac{k}{2}}}{\sqrt{N}}\sum^{N-1}_{n=\frac{N}{2}}e^{ikn}T^{n}S(1+T^{-2})|1,\frac{N}{2}+1\rangle\nonumber\\
&=&S(e^{ik}+e^{-ik})|\xi_{N/2-4}(k)\rangle+2S(e^{ik}+e^{-ik})\frac{|\xi_{N/2}(k)\rangle}{\sqrt{2}}\nonumber\\
&=&2S\cos k(|\xi_{N/2-4}(k)\rangle+\sqrt{2}|\xi_{N/2}(k)\rangle).\nonumber
\end{eqnarray}
\begin{figure}
\includegraphics[width=0.51\textwidth]{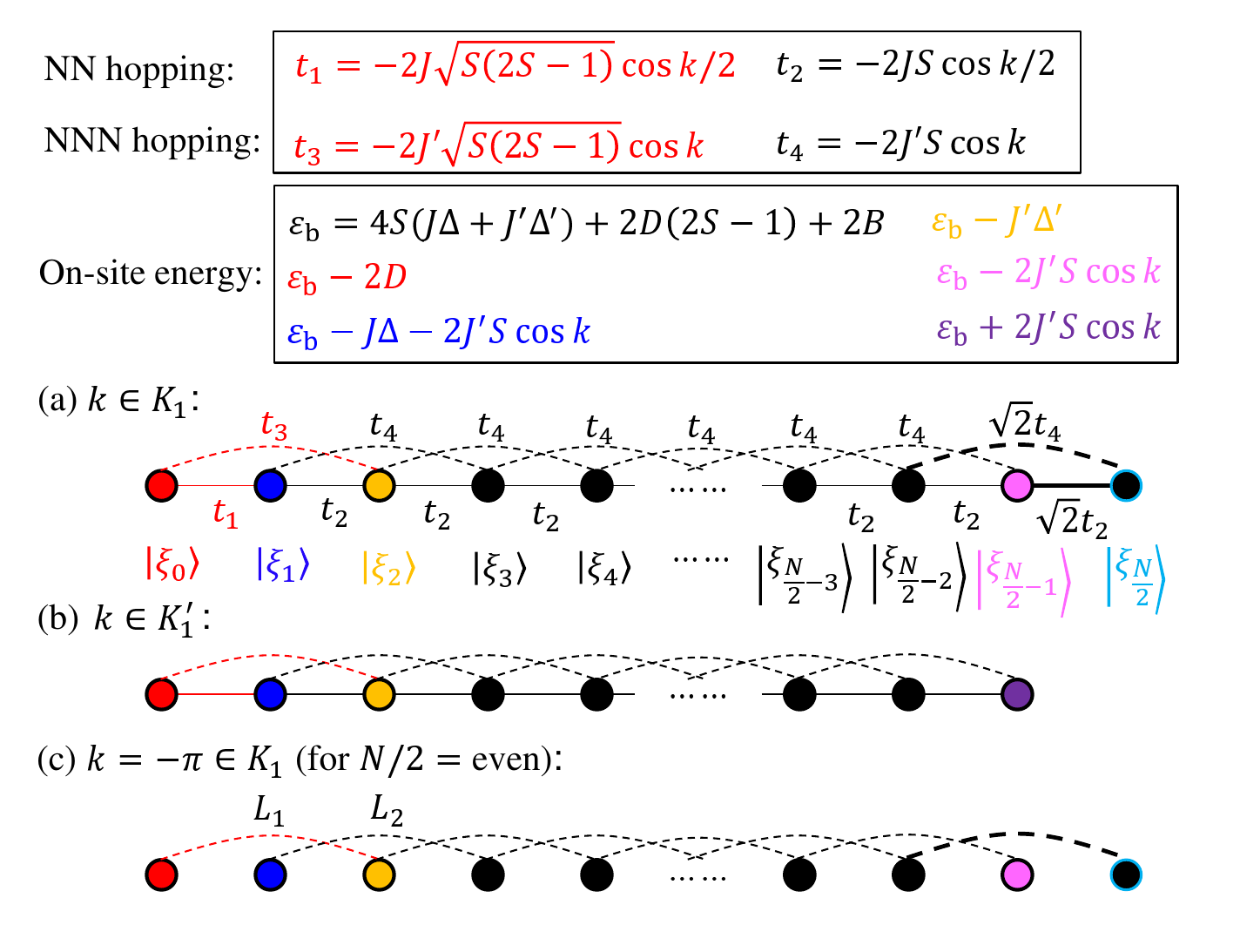}
\caption{(a) The matrix form of $H-E_F$ in the ordered basis $\{|\xi_0(k),\cdots,\xi_{N/2}(k)\}$ with $k\in K_1$ is represented by an effective lattice consisting of an open chain with both NN and NNN hoppings. The on-site energies and hopping strengths are indicated in respective colors. (b) The effective lattice for $k\in K'_1$.  (c) For $N$ divisible by 4, the special mode $k=-\pi$ lies in the set $K_1$, giving two decoupled open NN chains $L_1$ and $L_2$ of lengths $N/4+1$ and $N/4$, respectively.}
\label{Fig1}
\end{figure}
We see that $H_{\mathrm{NNN}}$ connects $|\xi_{N/2-2}(k)\rangle$ with $|\xi_{N/2-4}(k)\rangle$ and $|\xi_{N/2}(k)\rangle$. The remaining non-vanishing matrix elements of $\tilde{H}_{\mathrm{XX,NNN}}$ can be obtained in a similar way.
\par By gathering all the terms in $H$, we find that the matrix representation of $H$ in the basis can be represented by an effective lattice consisting of an open chain with both NN and NNN hoppings [Fig.~\ref{Fig1}(a)]. The action of $H-E_F$ on an arbitrary Bloch state can directly be read off from the lattice. For example, $(H-E_F)|\xi_1(k)\rangle=(\varepsilon_{\mathrm{b}}-J\Delta-2J'S\cos k)|\xi_1(k)\rangle+t_1|\xi_0(k)\rangle+t_2|\xi_2(k)\rangle+t_4|\xi_3(k)\rangle$, where $\varepsilon_{\mathrm{b}}=4S(J\Delta+J'\Delta')+2D(2S-1)+2B$, $t_1=-2J\sqrt{S(2S-1)}\cos\frac{k}{2}$, $t_2=-2JS\cos\frac{k}{2}$, and $t_4=-2J'S\cos k$. Note that the on-site energies on sites $|\xi_1(k)\rangle$ and $|\xi_{N/2-1}(k)\rangle$ are $k$-dependent. For $k\in K'_1$, the effective lattice can simply be obtained by removing the last site of the lattice since $|\xi_{N/2}(k)\rangle$ is not properly defined [Fig.~\ref{Fig1}(b)]. We thus convert the two-magnon problem into a single-particle one on an open chain. The two-magnon excitation energies $\mathcal{E}_2(k)$ as functions of $k$ can be obtained by diagonalizing the above matrices,
\begin{eqnarray}
(H-E_F)|\psi_{\alpha}(k)\rangle=\mathcal{E}_{2,\alpha}(k)|\psi_{\alpha}(k)\rangle,
\end{eqnarray}
where $\alpha=1,2,\cdots,N/2+1$ for $k\in K_1$ and $\alpha=1,2,\cdots,N/2$ for $k\in K'_1$.
\par In practice, solving the inhomogeneous open chain with NNN hopping involves the diagonalization of pentadiagonal matrices, which in general does not admit analytical solutions. We thus numerically diagonalize the effective chains to obtain the two-magnon excitations for systems of hundreds of spins.
\par However, for the special mode $k=-\pi$ the problem becomes, at least semianalytically, tractable. Actually, for $k=-\pi$ the NN hopping proportional to $\cos k/2$ vanishes and the effective open chain is separated into two decoupled NN chains $L_1$ and $L_2$ formed by $\{|\xi_0(-\pi)\rangle,|\xi_2(-\pi)\rangle,\cdots,|\xi_{N/2}(-\pi)\rangle\}$ and $\{|\xi_1(-\pi)\rangle,|\xi_3(-\pi)\rangle,\cdots,|\xi_{N/2-1}(-\pi)\rangle\}$, respectively [Fig.~\ref{Fig1}(c)]. For even $N/2$ with $k=-\pi\in K_1$, the effective Hamiltonians for $L_1$ and $L_2$ can both be incorporated into an inhomogeneous tridiagonal matrix
\begin{eqnarray}\label{hnn}
(h)_{(n+1)\times (n+1)}=\left(
      \begin{array}{cccccccc}
        a_1 &  b_1 &   & &    &    &   &   \\
         b_1 & a_2 &  b &  &   &   &   &   \\
          &  b & 0 &  b  & &   &   &   \\
          &   &  b &  0 &  &    &   &   \\
           &   &    &  & \ddots&   &   &   \\
          &   &   &     & &0 &  b &   \\
          &   &   &    &  &b & 0 &   b_2 \\
          &   &   &    & &  &  b_2  & a_3 \\
      \end{array}
    \right).
\end{eqnarray}
For example,
\par i) $L_1$ with $S>1/2$:
\begin{eqnarray}\label{SL1}
n&=&\frac{N}{4},~a_1=-2D,~a_2=-J'\Delta',~a_3=0,\nonumber\\
b_1&=&t_3,~b=t_4,~b_2=\sqrt{2}t_4.
\end{eqnarray}
\par ii) $L_1$ with $S=1/2$:
\begin{eqnarray}\label{halfL1}
n&=&\frac{N}{4}-1,~a_1=-J'\Delta',~a_2=a_3=0,\nonumber\\
b_1&=&b=t_4,~b_2=\sqrt{2}t_4.
\end{eqnarray}
\par iii) $L_1$ with $S\geq 1/2$:
\begin{eqnarray}\label{SL2}
n&=&\frac{N}{4}-1,~a_1=-J\Delta+2J'S,~a_2=0,~a_3=2J'S,\nonumber\\
b_1&=&b=b_2=t_4.
\end{eqnarray}
\subsection{The plane-wave ansatz}
\par We now provide a plane-wave ansatz solution~\cite{Grim,PRB2017,PRB2019} to the eigenvalue problem of the matrix $(h)_{(n+1)\times (n+1)}$ given by Eq.~(\ref{hnn}). Let
\begin{eqnarray}
(h)_{(n+1)\times (n+1)}\vec{v}=\lambda \vec{v},
\end{eqnarray}
where $\lambda$ and $\vec{v}=(v_1,\cdots,v_{n+1})^T$ are, respectively, the eigenvalue and eigenvector to be solved. Explicitly, we have four boundary equations
\begin{eqnarray}\label{BDeq}
a_1v_1+b_1v_2&=&\lambda v_1,\nonumber\\
b_1v_1+a_2v_2+bv_3&=&\lambda v_2,\nonumber\\
bv_{n-1}+b_2v_{n+1}&=&\lambda v_{n},\nonumber\\
b_2 v_{n}+a_3v_{n+1}&=&\lambda v_{n+1},
\end{eqnarray}
and $n-3$ bulk equations
\begin{eqnarray}\label{Bulkeq}
bv_{j-1}+bv_{j+1}=\lambda v_j,~j=3,4,\cdots,n-1.
\end{eqnarray}
The plane-wave ansatz assumes that
\begin{eqnarray}\label{planewave}
v_j=Xe^{ipj}+Ye^{-ipj},~j=2,3,\cdots,n
\end{eqnarray}
where $X$ and $Y$ are $j$-independent coefficients to be determined. The end components of $\vec{v}$, $v_1$ and $v_{n+1}$, can be obtained from the first and the last boundary equations:
\begin{eqnarray}\label{vend_app}
v_1&=&\frac{b_1}{\lambda-a_1}v_2,~v_{n+1}=\frac{b_2}{\lambda-a_3}v_n.
\end{eqnarray}
The bulk equations simple give
\begin{eqnarray}\label{lambdacosp}
\lambda =2 b\cos p.
\end{eqnarray}
To determine the allowed values of the wave number $p$, we apply the ansatz in the four boundary equations. After eliminating $v_1$ and $v_{n+1}$~\cite{PC}, we get
\begin{eqnarray}\label{c1c2M}
\left(
  \begin{array}{cc}
    c^{(1)}_p & c^{(1)}_{-p} \\
    c^{(2)}_p & c^{(2)}_{-p} \\
  \end{array}
\right)\left(
         \begin{array}{c}
           X \\
           Y \\
         \end{array}
       \right)=0,
\end{eqnarray}
where
\begin{eqnarray}\label{c1c2}
c^{(1)}_p&=&b^2-(a_1+a_2)be^{ip} +(b^2+a_1a_2-b^2_1)e^{i2p}-a_2b e^{i3p},\nonumber\\
c^{(2)}_p&=&e^{inp}[(1+e^{i2p})b^2-b^2_2-a_3be^{ip}].
\end{eqnarray}
To obtain nontrivial solutions of $(X,Y)$, the determinant of the $2\times 2$ matrix appearing in the above equation must vanish, i.e., $c^{(1)}_pc^{(2)}_{-p}-c^{(2)}_pc^{(1)}_{-p}=0$, which after some manipulation becomes
\begin{widetext}
\begin{eqnarray}\label{Treq}
\frac{\tan np}{\sin p}=\frac{(a_1a_2-b^2_1-2a_2b\cos p)[a_3b+(b^2_2-2b^2)\cos p]-b^2_2[a_1b+2a_2b\cos^2p-(a_1a_2+2b^2-b^2_1)\cos p]}{[a_3b+(b^2_2-2b^2)\cos p][(a_1a_2+2b^2-b^2_1)\cos p-a_1b-2a_2 b\cos^2 p]+b^2_2(b^2_1+2a_2b\cos p-a_1a_2)\sin^2p}.
\end{eqnarray}
\end{widetext}
It is apparent that if $p$ is a solution of the above equation, so is $2\pi-p$. We thus need only to solve the above equation on the interval $p\in [0,\pi]$. However, it is possible that the number of real solutions of Eq.~(\ref{Treq}) is less than $n+1$. In this case, one has to pursue complex solutions of Eq.~(\ref{Treq}).
\par For each allowed $p$, Eqs.~(\ref{c1c2M}) and (\ref{c1c2}) lead to the following (unnormalized) wave functions (for $j=2,3,\cdots,n$)
\begin{eqnarray}\label{vj_p}
v_j&=&e^{ip(j-n-1)}[e^{ip}(b^2-b^2_2)+e^{-ip}b^2-a_3b]\nonumber\\
&&-e^{-ip(j-n-1)}[e^{-ip}(b^2-b^2_2)+e^{ip}b^2-a_3b].\nonumber\\
\end{eqnarray}
The method and results presented in this subsection will used to solve the open chains $L_1$ and $L_2$ for $k=-\pi$.
\section{Exact zero-excitation-energy states when $D=0$}\label{SecZEES}
\par Before discussing the two-magnon excitations in detail, let us first study a related problem, i.e., the existence of zero-excitation-energy states (ZEESs) (with respect to the ferromagnetic state $|F\rangle$) under certain conditions. It is shown in Ref.~\cite{PRB2022} that for the spin-$S$ XXZ chain in the absence of the SI term and the magnetic field ($J'=D=B=0$), if the condition $\Delta=\cos k,~k\in K_0$ is satisfied, there then exists a series of ZEESs,
\begin{eqnarray}
(H-E_F)|_{J'=D=B=0}(L_k)^n|F\rangle=0,~n\leq 2NS,
\end{eqnarray}
where $L_k=\sum^N_{j=1}e^{ikj}S^-_j$ is a collective spin lowering operator. The (unnormalized) ZEES $(L_k)^n|F\rangle$ carries momentum $nk~(\mathrm{mod}~2\pi)$ since $T(L_k)^n|F\rangle=e^{-ink}(L_k)^n|F\rangle$. In this section, we explore the condition for the existence of ZEESs for the $J-J'$ chain $H|_{D=B=0}=H_{\mathrm{NN}}+H_{\mathrm{NNN}}$.
\subsection{Condition for the existence of zero-excitation-energy states}
\par We first look at the simplest case of $n=1$. It is easy to check that
\begin{eqnarray}
&&~[H|_{D=B=0},L_k]\nonumber\\
&=&J\sum^N_{n=1}e^{ikn}[(\Delta e^{ik}-1)S^-_{n+1}S^z_n +(\Delta-e^{ik})S^-_{n}S^z_{n+1}]\nonumber\\
&+&J'\sum^N_{n=1}e^{ikn}[(\Delta' e^{i2k}-1)S^-_{n+2}S^z_n +(\Delta'-e^{i2k})S^-_{n}S^z_{n+2}],\nonumber\\
\end{eqnarray}
which gives
\begin{eqnarray}
&&(H-E_F)|_{D=B=0}L_k|F\rangle\nonumber\\
&=&2S[J(\Delta-\cos k)+J'(\Delta'-\cos 2k)]L_k|F\rangle.
\end{eqnarray}
The one-magnon state $L_k|F\rangle$ is thus a ZEES when
\begin{eqnarray}\label{cond1}
J(\Delta-\cos k)+J'(\Delta'-\cos 2k)=0,~k\in K_0
\end{eqnarray}
is fulfilled. This is reasonable since the left-hand side of the above equation is proportional to the one-magnon excitation energy $\mathcal{E}_1(k)|_{D=B=0}$ given by Eq.~(\ref{E1magnonBloch}). To see whether $L^2_k|F\rangle$ is a two-magnon ZEES under the above condition, we further calculate
\par
\begin{eqnarray}
&&~[L_k,[H|_{D=B=0},L_k]]\nonumber\\
&=&2J\sum^N_{n=1}e^{i(2n+1)}(\Delta-\cos k)S^-_{n}S^-_{n+1}\nonumber\\
&+&2J'\sum^N_{n=1}e^{i2(n+1)k}(\Delta'-\cos 2k) S^-_{n}S^-_{n+2}.
\end{eqnarray}
By applying both sides of the above equation to $|F\rangle$, we see that Eq.~(\ref{cond1}) is not a sufficient condition for $L^2_k|F\rangle$ being a ZEES. We must impose a stronger condition
\begin{eqnarray}\label{cond2}
\Delta-\cos k~\mathrm{and}~\Delta'-\cos 2k,~k\in K_0
\end{eqnarray}
to guarantee $[L_k,[H|_{D=B=0},L_k]]=0$, and hence $(H-E_F)|_{D=B=0}L^2_k|F\rangle=0$.
\par Note now that $[L_k,[L_k,[H|_{D=B=0},L_k]]]=0$, $[L_k,[L_k,[L_k,[H|_{D=B=0},L_k]]]]=0$, $\cdots$ are always true if $[L_k,[H|_{D=B=0},L_k]]=0$, we immediately get
\begin{eqnarray}
(H-E_F)|_{D=B=0}(L_k)^n|F\rangle=0,~n\leq 2NS,
\end{eqnarray}
under the condition (\ref{cond2}).
\par A direct consequence of the above analysis is that, under the condition given by (\ref{cond2}), the lowest $n$-magnon excitation energy, $\mathcal{E}^{(\min)}_n(k)$, must be nonpositive.
\par We now explicitly show that, for any $k\in K_0$ and $|k|\leq\pi/2$ (such that $2k$ lies in the first Brillouin zone), the following two-magnon state (in the ordered basis $\{|\xi_0(2k)\rangle, |\xi_1(2k)\rangle,\cdots,|\xi_{N/2}(2k)\rangle\}$)~\cite{PRB2022}
\begin{eqnarray}\label{PsiZEES}
|\Psi_{\mathrm{ZEES}}\rangle=\left(\tilde{S}/2S,1,\cdots,1,1/\sqrt{2}\right)^T
\end{eqnarray}
where $\tilde{S}\equiv\sqrt{S(2S-1)}$, is a ZEES under the condition given by Eq.~(\ref{cond2}). Actually, for $\Delta=\cos k$ and $\Delta'=\cos 2k$ the matrix form of the effective lattice representing $H-E_F$ reads
\begin{widetext}
\begin{eqnarray}
4S(\mathcal{J}_k+\mathcal{J}'_k)-\left(
  \begin{array}{cccccccccc}
    0 & 2\tilde{S}\mathcal{J}_k & 2\tilde{S}\mathcal{J}'_k &    &   &   &   &   &   &   \\
    2\tilde{S}\mathcal{J}_k & \mathcal{J}_k+2S\mathcal{J}'_k & 2S\mathcal{J}_k & 2S\mathcal{J}'_k &   &   &   &   &   &   \\
    2\tilde{S}\mathcal{J}'_k & 2S\mathcal{J}_k & \mathcal{J}'_k & 2S\mathcal{J}_k & 2S\mathcal{J}'_k &   &   &   &   &   \\
       &  2S\mathcal{J}'_k & 2S\mathcal{J}_k & 0& 2S\mathcal{J}_k & \ddots  &   &   &   \\
       &   & 2S\mathcal{J}'_k & 2S\mathcal{J}_k & 0 & \ddots  &  &  &  &  \\
      &  &  & \ddots & \ddots & \ddots &  &  &  &  \\
      &  &  &  &  &  & \ddots &  &  &   \\
      &  &  &  &  &  &  & 0 & 2S \mathcal{J}_k & 2\sqrt{2}S\mathcal{J}'_k \\
      &  &  &  &  &  &  & 2S\mathcal{J}_k & 2S\mathcal{J}'_k & 2\sqrt{2}S\mathcal{J}_k \\
      &  &  &  &  &  &  & 2\sqrt{2}S\mathcal{J}'_k & 2\sqrt{2}S\mathcal{J}_k & 0 \\
  \end{array}
\right),
\end{eqnarray}
\end{widetext}
where $\mathcal{J}_k\equiv J\cos k$ and $\mathcal{J}'_k\equiv J'\cos 2k$. It is easy to check that $|\Psi_{\mathrm{ZEES}}\rangle$ is an eigenvector of the above matrix with zero eigenvalue.
\subsection{The isotropic case: $\Delta=\Delta'=1$}
\par In the isotropic case of $\Delta=\Delta'=1$, the total angular momentum $\vec{S}_{\mathrm{tot}}=\sum_j\vec{S}_j$ is conserved. It is obvious that the condition (\ref{cond2}) is satisfied if and only if $k=0$. Thus, the lowest $n$-magnon excitation energy $\mathcal{E}^{(\min)}_n(k)$ is nonpositive. In particular, if the ferromagnetic state $|F\rangle$ is a ground state, we must have
\begin{eqnarray}\label{necess}
\mathcal{E}^{(\min)}_n(k)=\mathcal{E}_n(0)=0,~n=1,\cdots,2NS
\end{eqnarray}
Thus, all the $2NS$ states $(L_0)^n|F\rangle,~n=1,\cdots,2NS$ are degenerate with $|F\rangle$ and possess energy $E_F|_{D=B=0}=-NS^2(J+J')$, indicating that the ground state is at least $(2NS+1)$-fold degenerate. These $2NS+1$ states all have total angular momentum $NS$.
\par Equation~(\ref{necess}) gives the necessary conditions for the ferromagnetic ground state. We define $(J/|J'|)^{(n)}_{\mathrm{th}} (N)$ as the threshold above which Eq.~(\ref{necess}) is satisfied for $n$. We see from Eq.~(\ref{Rth}) that
\begin{eqnarray}
(J/|J'|)^{(1)}_{\mathrm{th}}=4\cos^2\frac{\pi}{N},
\end{eqnarray}
which is just the necessary condition obtained in Ref.~\cite{Bader1979} by considering one-magnon excitations. However, for $n>1$ the threshold $(J/|J'|)^{(n)}_{\mathrm{th}}$ can only be determined numerically. The sufficient condition for the ferromagnetic ground state is obviously $J/|J'|\geq (J/|J'|)^{(\mathrm{FM})}_{\mathrm{th}}$, where
\begin{eqnarray}
(J/|J'|)^{(\mathrm{FM})}_{\mathrm{th}}= \max\{(J/|J'|)^{(n)}_{\mathrm{th}}|n=1,2,\cdots,NS\}.
\end{eqnarray}
\begin{figure}
\includegraphics[width=0.51\textwidth]{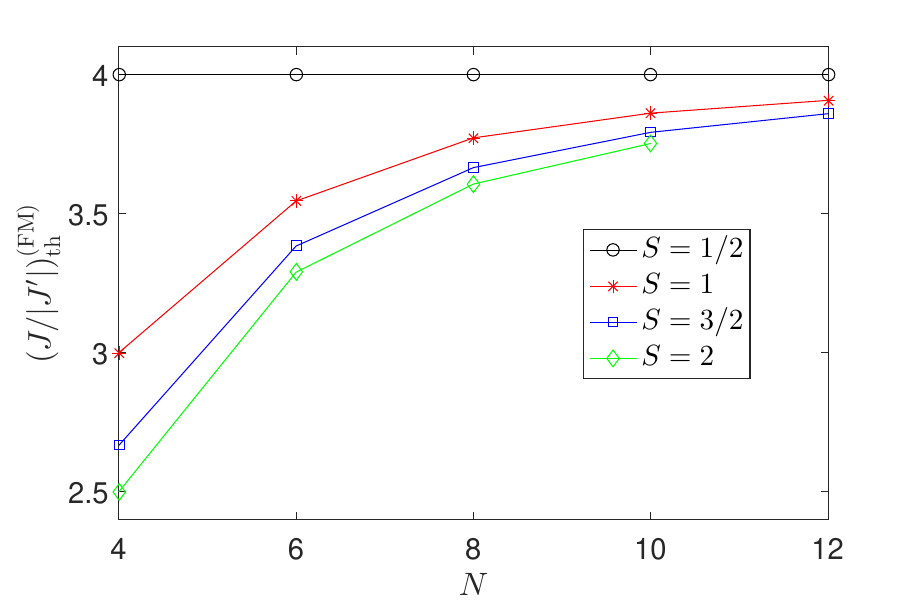}
\caption{The threshold $(J/|J'|)^{(\mathrm{FM})}_{\mathrm{th}}$ above which the ground state of the isotropic $J-J'$ chain with $\Delta=\Delta'=1$ is ferromagnetic. Other parameters: $B=D=0$.}
\label{thFM}
\end{figure}
\par Figure~\ref{thFM} shows $(J/|J'|)^{(\mathrm{FM})}_{\mathrm{th}}$ for different $S$ and $N$ obtained by exact diagonalization. For $S=1/2$, $(J/|J'|)^{(\mathrm{FM})}_{\mathrm{th}}$ is independent of $N$ and is always 4. Actually, Hamada, Kane, Nakagawa, and Natsume showed that at the point $J/|J'|= 4$ the ground state for $S=1/2$ is $(N+2)$-fold degenerate: besides the above-mentioned $N+1$ ferromagnetic states with total angular momentum $N/2$, there exists an additional state with zero total angular momentum that can be expressed as a linear combination of singlet bonds uniformly distributed on all sites~\cite{Exact1988}.
\par We see from Fig.~\ref{thFM} that $(J/|J'|)^{(\mathrm{FM})}_{\mathrm{th}}$ shows size dependence for $S>1/2$. For $N=4$, the threshold is shown to be $(J/|J'|)^{(\mathrm{FM})}_{\mathrm{th}}(4)=2+1/S$~\cite{Bader1979}. As $N$ increases, $(J/|J'|)^{(\mathrm{FM})}_{\mathrm{th}}$ increases monotonically and we expect that
\begin{eqnarray}
\lim_{N\to \infty}(J/|J'|)^{(\mathrm{FM})}_{\mathrm{th}}= 4,~(S>1/2)
\end{eqnarray}
Actually, Bader and Schilling showed that for $J/|J'|\geq 4$ the ground state of $H|_{D=B=0,\Delta=\Delta'=1}$ is ferromagnetic for arbitrary $S$~\cite{Bader1979}. By noting that $\lim_{N\to\infty}(J/|J'|)^{(1)}_{\mathrm{th}}=4$, we also expect that
\begin{eqnarray}\label{JnTL}
\lim_{N\to \infty}(J/|J'|)^{(n)}_{\mathrm{th}}= 4,~n=1,\cdots,2NS.
\end{eqnarray}
The results obtained in this section will be found useful in the discussion of the two-magnon excitations below.
\section{Two-magnon excitations}\label{SecIV}
\par In this section, we will study the two-magnon excitations in the $J-J'$ chain in detail by using the Bloch Hamiltonians we constructed in Sec.~\ref{SecII}.
\subsection{$S=1/2$}\label{SecIVA}
\par To verify the validity of our formalism, let us first study the case of $S=1/2$, which has been extensively studied using various methods~\cite{Chubukov1991,Kuzian2007,Furusaki2007,Furusaki2008}. It has been observed in previous works that there is always a region in the momentum space (usually near the band edge $k=-\pi$) supporting multimagnon bound states~\cite{Furusaki2007}. We will analytically demonstrate this fact in the two-magnon sector. As mentioned earlier, we set $D=0$ for $S=1/2$.
\begin{figure}
\includegraphics[width=0.53\textwidth]{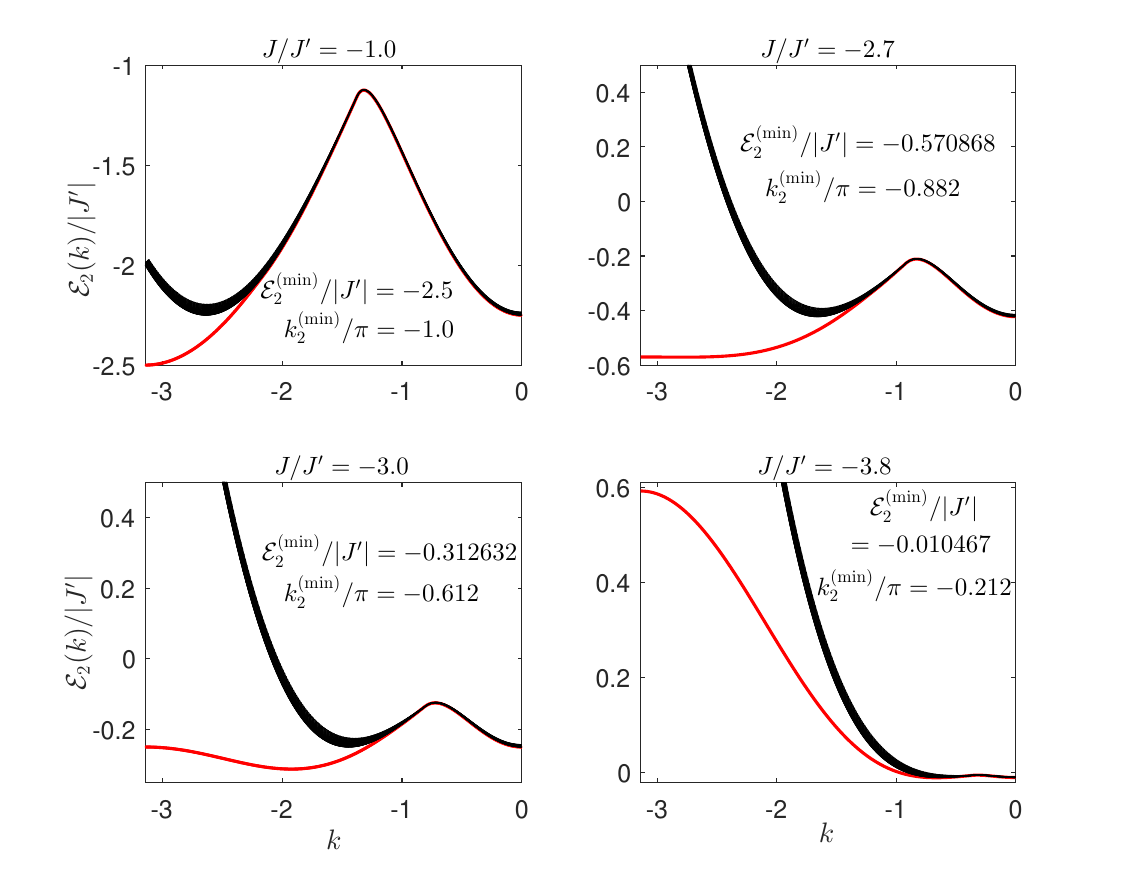}
\caption{The lowest 20 two-magnon excitation levels $\mathcal{E}_2(k)/|J'|$ for a spin-1/2 chain with $J/J'=-1.0,-2.7,-3.0$, and $-3.8$. The red curves indicate the lowest levels contributed by the two-magnon bound states. Other parameters: $N=1000$, $\Delta=\Delta'=1$, and $B=D=0$.}
\label{Jhalfkeck}
\end{figure}
\par Since $(S^-_j)^2=0$ for $S=1/2$, the leftmost site in Fig.~\ref{Fig1} is absent and $t_1=t_3=0$. Figure~\ref{Jhalfkeck} shows the lowest twenty two-magnon excitation levels $\mathcal{E}_2(k)/|J'|$ on $k\in[-\pi,0]$. We choose $N=1000$, $\Delta=\Delta'=1$, $B=0$, and $J/J'=-1.0, -2.7, -3.0, -3.8$, in accordance with Ref.~\cite{Furusaki2007}. We see that our exact results for a finite-size system agree well with that obtained in Ref.~\cite{Furusaki2007} for infinite systems (note that certain truncations of the Hilbert space were adopted there): for $-4<J/J'<0$ (so that the ground state is not ferromagnetic) there always exists a region in the momentum space where the two-magnon bound states are the lowest ones with negative excitation energies.
\begin{figure}
\includegraphics[width=0.52\textwidth]{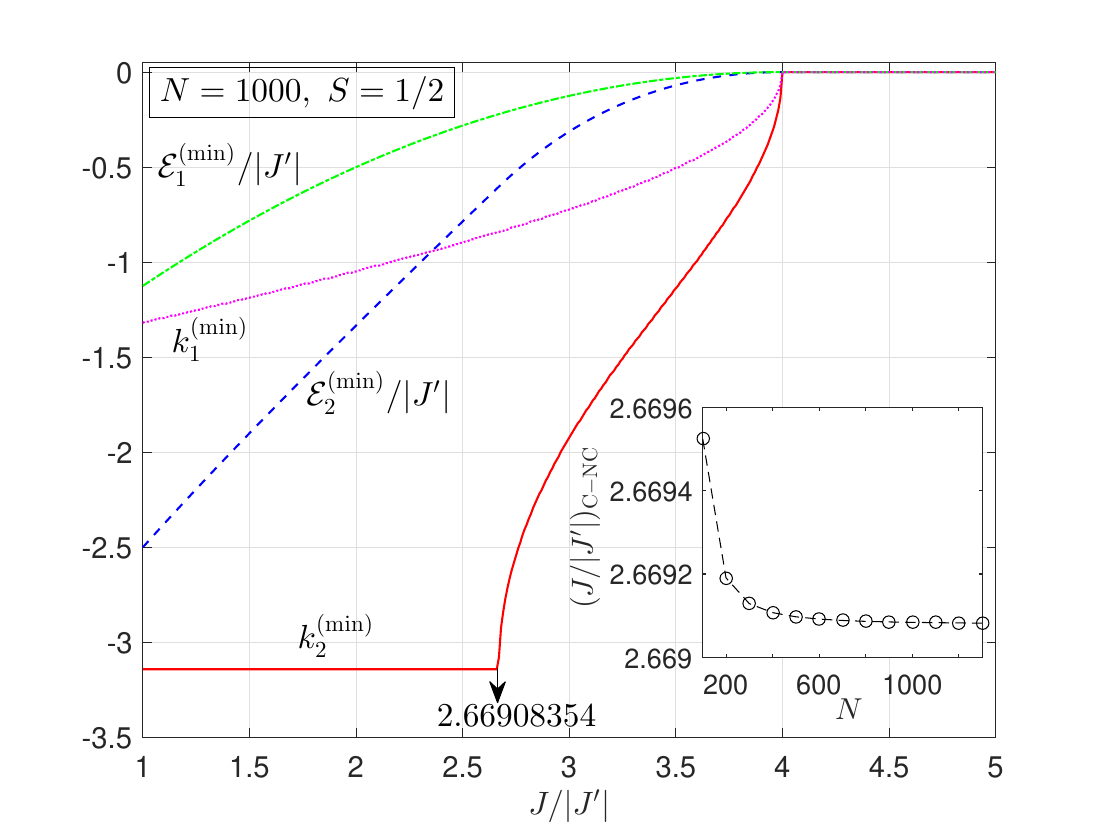}
\caption{Evolution of the lowest two-magnon excitation energy $\mathcal{E}^{(\min)}_2/|J'|=\mathcal{E}_2(k^{(\min)}_2)/|J'|$ (blue dashed) and the corresponding wave number $k^{(\min)}_2$ (red solid) with increasing $J/|J'|$ for $N=1000$ and $S=1/2$. Also shown is the minimal one-magnon excitation energy $\mathcal{E}^{(\min)}_1$ (green dash-dotted) and the corresponding $k^{(\min)}_1$ (pink dotted). The inset shows the value of $(J/|J'|)_{\mathrm{C-NC}}$ with increasing number of sites $N$ ($100$ to $1300$). Other parameters: $\Delta=\Delta'=1$ and $B=D=0$.}
\label{twomagnonEminKmin}
\end{figure}
\par Figure~\ref{twomagnonEminKmin} shows the lowest two-magnon excitation energy $\mathcal{E}^{(\min)}_2/|J'|=\mathcal{E}_2(k^{(\min)}_2)/|J'|$ (blue dashed curve) as a function of $J/|J'|$, where $k^{(\min)}_2$ (red solid curve) is the mode corresponding to this minimum excitation. For $J/|J'|\geq 4$, the ground state is ferromagnetic and highly degenerate~\cite{Bader1979,Exact1988}. According to the analysis in Sec.~\ref{SecZEES}, the lowest two-magnon eigenstate is the ZEES $(L_0)^2|F\rangle\sim (1,1,\cdots,1/\sqrt{2})^T$ (in the basis $\{|\xi_1(0)\rangle,|\xi_2(0)\rangle,\cdots,|\xi_{N/2}(0)\rangle\}$), which explains $k^{(\min)}_2=0$ and $\mathcal{E}^{(\min)}_2/|J'|=0$ in this regime.
\par For $0<J/|J'|<4$, $\mathcal{E}^{(\min)}_2/|J'|$ is negative but increases with increasing $J/|J'|$. Meanwhile, $k^{(\min)}_2$ is no longer zero and there exists a so-called commensurate-incommensurate (C-NC) transition below which one has $k^{(\min)}_2=-\pi$. Our numerical result fixes the C-NC transition to be $(J/J')_{\mathrm{C-NC}}=-2.66908354$ for $N=1000$, which is very close to the result obtained from Green's function analysis for an infinite chain (i.e., $1/0.37466105983527\approx-2.66907909$)~\cite{Kuzian2007}.  The inset of Fig.~\ref{twomagnonEminKmin} shows the size dependence (up to $N=1300$) of the C-NC transition point, showing that $(J/J')_{\mathrm{C-NC}}$ decreases asymptotically with $N$ and approaches the value in the thermodynamic limit as $N\to\infty$. For completeness, we also plot the lowest one-magnon excitation energy $\mathcal{E}^{(\min)}_1/|J'|$ and the corresponding $k^{(\min)}_1$. It can be seen that $\mathcal{E}^{(\min)}_1>\mathcal{E}^{(\min)}_2$ for $0\leq J/|J'|<4$~\cite{Chubukov1991,Dmitriev2006,Vekua2006,Vekua2007,Kuzian2007}.
\par Below $(J/J')_{\mathrm{C-NC}}$ the lowest two-magnon excitation has a commensurate momentum $k^{(\min)}_2=-\pi$, which deserves further investigation. As mentioned in Sec.~\ref{SecII}, both the two decoupled chains $L_1$ and $L_2$ can be solved through the plane-wave ansatz~\cite{Grim,PRB2017,PRB2019}.
\par Let the eigevectors for $L_\alpha$ ($\alpha=1,2$, note that the site $|\xi_0(-\pi)\rangle$ in $L_1$ is absent when $S=1/2$) be
\begin{eqnarray}
V^{(\alpha)}=(V^{(\alpha)}_1,\cdots,V^{(\alpha)}_{N/4})^T,
\end{eqnarray}
with
\begin{eqnarray}
V^{(\alpha)}_j=X_\alpha e^{ip_\alpha j}+Y_\alpha e^{-ip_\alpha j},~j=2,\cdots,\frac{N}{4}-1
\end{eqnarray}
where $X_\alpha$ and $Y_\alpha$ are two $j$-independent coefficients and $p_\alpha$ is a wave number to be determined. From Eq.~(\ref{lambdacosp}), the excitation energies (eigenenergies of $H-E_F$) are given by
\begin{eqnarray}\label{Erealp}
E^{(\alpha)}(p_\alpha)=2(J\Delta+B)+2J'(\Delta'+\cos p_\alpha).
\end{eqnarray}
According to the correspondence in Eq.~(\ref{halfL1}), the wave number $p_1$ satisfies the transcendental equation
\begin{eqnarray}\label{Teq1}
\tan \left(\frac{N}{4}-1\right)p_1& =&f_1(\Delta',p_1),\nonumber\\
f_1(\Delta',p_1)&\equiv&\frac{\cos p_1+\Delta'}{\sin p_1}.
\end{eqnarray}
Since the correspondence given by Eq.~(\ref{SL2}) is valid for $S\geq 1/2$, we have, for both $S=1/2$ and $S>1/2$,
\begin{eqnarray}\label{Teq2}
\frac{\tan\left(\frac{N}{4}-1\right)p_2}{\sin p_2}&=&f_2(\tilde{j},p_2),\nonumber\\
f_2(\tilde{j},p_2)&\equiv&\frac{\tilde{j} -4S(1-\cos p_2)}{(1-\cos p_2)(4S\cos p_2+\tilde{j})},
\end{eqnarray}
where we defined $\tilde{j}\equiv J\Delta/J'<0$.
\par Equations (\ref{Teq1}) and (\ref{Teq2}) have to be solved on the interval $p_\alpha\in[0,\pi]$, $\alpha=1,2$. In general, these equations do not admit analytical solutions. However, they can be solved graphically by plotting both sides of the equation as functions of $p_\alpha$.
\begin{figure}
\includegraphics[width=0.53\textwidth]{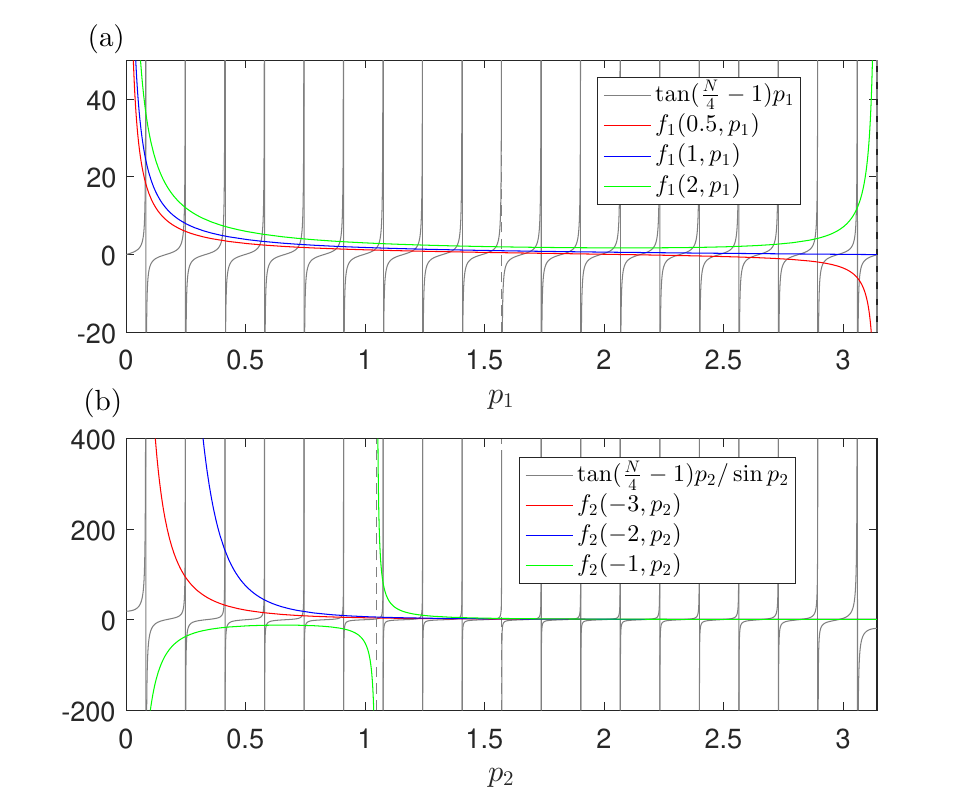}
\caption{(a) The functions $\tan \left(\frac{N}{4}-1\right)p_1$ (gray) and $f_1(\Delta',p_1)$ appearing in Eq.~(\ref{Teq1}) for $\Delta'=0.5$ (red), $1$ (blue), and $2$ (green). (b) The functions $\tan \left(\frac{N}{4}-1\right)p_2/\sin p_2$ (gray) and $f_2(\tilde{j},p_1)$ (with $\tilde{j}=J\Delta/J'$) appearing in Eq.~(\ref{Teq2}) for $S=1/2$ and $\tilde{j}=-3$ (red), $-2$ (blue), $-1$ (green). Here, we choose $N=80$. }
\label{N40tan}
\end{figure}
\par i) Solution of $L_1$ [for $S=1/2$,~Eq.~(\ref{Teq1})].
\par Note that the function $\tan \left(\frac{N}{4}-1\right)p_1$ diverges at $p_1=\frac{2\pi}{N-4}, \frac{6\pi}{N-4},\cdots,\frac{(N-6)\pi}{N-4}$, dividing the interval $[0,\pi]$ into $N/4$ ones [see Fig.~\ref{N40tan}(a)]:
\begin{eqnarray}\label{Intervals}
\left[0,\frac{2\pi}{N-4}\right],~\left[\frac{2\pi}{N-4},\frac{6\pi}{N-4}\right],~\cdots,\left[\frac{(N-6)\pi}{N-4},\pi\right].\nonumber
\end{eqnarray}
The first and last intervals will be respectively denoted as $I_{\mathrm{L}}=\left[0,\frac{2\pi}{N-4}\right]$ and $I_{\mathrm{R}}=\left[\frac{(N-6)\pi}{N-4},\pi\right]$. Note that $\tan \left(\frac{N}{4}-1\right)p_1\geq 0$ ($\leq 0$) on $I_{\mathrm{L}}$ ($I_{\mathrm{R}}$).
\par It is obvious that the solutions of Eq.~(\ref{Teq1}) are independent of $J/|J'|$ and determined only by the value of $\Delta'$. Below we consider three different cases.
\par i-a) $0<\Delta'<1$.
\par We have
\begin{eqnarray}
\lim_{p_1\to0^+}f_1(\Delta',p_1)=+\infty,~\lim_{p_1\to\pi^-}f_1(\Delta',p_1)=-\infty,
\end{eqnarray}
giving exactly $N/4$ real solutions, each of which lies in one of the above $N/4$ intervals [Fig.~\ref{N40tan}(a), red curve].
\par i-b) $\Delta'=1$.
\par There are still $N/4$ real solution, including an obvious one, $p_1=\pi$ [Fig.~\ref{N40tan}(a), blue curve], which gives the highest excitation energy
\begin{eqnarray}\label{Emaxppi}
E^{(1)}(\pi)|_{\Delta'=1}=2(J\Delta+B).
\end{eqnarray}
For this special solution, the plane-wave ansatz does not work since the $c^{(1)}_{\pi}$ and $c^{(2)}_{\pi}$ given by Eq.~(\ref{c1c2}) are both zero. However, it is easy to check that the vector
\begin{eqnarray}\label{Vmaxppi}
V^{(1)}(\pi)|_{\Delta'=1}=\frac{2}{\sqrt{N-2}}(1,-1,\cdots,1,-1,1,-1/\sqrt{2})^T\nonumber
\end{eqnarray}
solves the eigenvalue problem, indicating that the highest excited state is indeed an extended state.
\par i-c) $\Delta'>1$.
\par We have
\begin{eqnarray}
\lim_{p_1\to\pi^-}f_1(\Delta',p_1)=+\infty,
\end{eqnarray}
which gives $N/4-1$ real solutions since no intersections exist on $I_{\mathrm{R}}$ [Fig.~\ref{N40tan}(a), green curve].
\par Actually, there exists a complex solution
\begin{eqnarray}
p_1=m\pi+i\tilde{p}_1
\end{eqnarray}
for $\Delta'>1$, where $m$ is an integer (to ensure the reality of the eigenvalue) and $\tilde{p}_1$ is real~\cite{PRB2017,PRB2019}. Accordingly, equation (\ref{Teq1}) becomes
\begin{eqnarray}\label{Teq1tanh}
 \tanh \left(\frac{N}{4}-1\right)\tilde{p}_1& =& -\frac{\cosh \tilde{p}_1+(-1)^m\Delta'}{ \sinh \tilde{p}_1}.
\end{eqnarray}
The above equation has to be solved on $\tilde{p}_1\in(0,\infty)$. It is easy to see that it has no solutions unless $m$ is odd (the right-hand side of the above equation is always negative for even $m$). In addition, there is no solution near $\tilde{p}_1=0$ since the right-hand side diverges as $\tilde{p}_1\to 0^+$. As a result, we have $\tanh \left(\frac{N}{4}-1\right)\tilde{p}_1\approx 1$ for large enough $N$, giving
\begin{eqnarray}
\tilde{p}_1\approx \ln\Delta'>0.
\end{eqnarray}
The corresponding eigenenergy
\begin{eqnarray}\label{Ecompppi}
E^{(1)}_{\mathrm{NNN-Ex}}&=&2(J\Delta+B)+2J'(\Delta'-\cosh\tilde{p}_1)\nonumber\\
&\approx& 2(J\Delta+B)+J'(\Delta'-1/\Delta')
\end{eqnarray}
is the highest level for $J'<0$. To see the nature of this highest state, we obtain from Eqs.~(\ref{halfL1}) and (\ref{vj_p}) the bulk components of the wave function,
\begin{eqnarray}\label{Vcompppi}
V^{(1)}_{\mathrm{NNN-Ex},j}=(-1)^j \cosh[(N/4-j)\tilde{p}_1],
\end{eqnarray}
with $j=2,\cdots,N/4-1$. The two end components can be obtained from Eq.~(\ref{vend_app}) as
\begin{eqnarray}\label{Vcompppi_end}
V^{(1)}_{\mathrm{NNN-Ex},1}&=&-\Delta'V^{(1)}_{\mathrm{NNN-Ex},2},\nonumber\\
V^{(1)}_{\mathrm{NNN-Ex},N/4}&=&-\frac{\sqrt{2}}{\Delta'+1/\Delta'}V^{(1)}_{\mathrm{NNN-Ex},N/4-1}.
\end{eqnarray}
It is apparent that (for $\Delta'>1$)
\begin{eqnarray}
|V^{(1)}_{\mathrm{NNN-Ex},1}|>|V^{(1)}_{\mathrm{NNN-Ex},2}|>\cdots>|V^{(1)}_{\mathrm{NNN-Ex},N/4}|,
\end{eqnarray}
indicating that the state is localized around the left end of the $L_1$ chain (i.e., the site $|\xi_2(-\pi)\rangle$) and corresponds to a two-magnon bound state with the two spin deviations being mainly located on two NNN sites in real space. This state will be referred to as a next-nearest-neighbor exchange (NNN-Ex) two-magnon bound state below.
\par ii) Solution of $L_2$ [for $S\geq 1/2$, Eq.~(\ref{Teq2})].
\par The solutions of Eq.~(\ref{Teq2}) depend only on the value of $\tilde{j}=J\Delta/J'$. We will prove the following
\par \emph{Proposition}. For any $S\geq 1/2$ and for all $J/J'<0$ and $\Delta>0$, there always exists a two-magnon bound state below the scattering continuum in the $k=-\pi$ sector. For $S=1/2$, this state is just the lowest two-magnon excited state.
\par \emph{Proof}. We first show that, for all $J/J'<0$ and $\Delta>0$, Eq.~(\ref{Teq2}) has exactly $N/4-1$ real solutions on $p_2\in(0,\pi)$. The function $\tan(\frac{N}{4}-1)p_2/\sin p_2$ has the same set of singularities as $\tan(\frac{N}{4}-1)p_2$ on $p_2\in(0,\pi)$ and is positive (negative) on on $I_{\mathrm{L}}$ ($I_{\mathrm{R}}$). Note that the numerator of $f_2(\tilde{j},p_2)$ is always negative, we have [here, $\mathrm{sgn}(x)\equiv x/|x|$]
\begin{eqnarray}
\lim_{p_2\to 0^+}f_2(\tilde{j},p_2)&=&-\mathrm{sgn}(4S\cos p_2+\tilde{j})\infty,\nonumber\\
\lim_{p_2\to \pi^-}f_2(\tilde{j},p_2)&=&\frac{1}{2}+\frac{2S}{4S-\tilde{j}}>0.
\end{eqnarray}
We also need to know the behavior the derivative of $f_2(\tilde{j},p_2)$. Let $\tilde{c}\equiv\cos p_2\in[-1,1)$, we get
\begin{eqnarray}\label{partialf2}
\partial_{\tilde{c}}f_2(\tilde{j},\tilde{c})=\frac{16S^2\tilde{c}^2+8S(\tilde{j}-4S)\tilde{c}+\tilde{j}^2-4S\tilde{j}+16S^2}{(\tilde{c}-1)^2(\tilde{j}+4S\tilde{c})^2}.\nonumber\\
\end{eqnarray}
\par We consider two different cases.
\par i) $\tilde{j}\leq -4S$.
\par In this case, $f_2(\tilde{j},p_2)$ is a positive regular function on $p_2\in(0,\pi)$ and $\lim_{p_2\to 0^+}f_2(\tilde{j},p_2)=+\infty$. As a quadratic function of $\tilde{c}$, the numerator in Eq.~(\ref{partialf2}) is $\tilde{j}(\tilde{j}+4S)\geq0$ at $\tilde{c}=1$ and is $(\tilde{j}-6S)^2+28S^2>0$ at $\tilde{c}=-1$, and the axis of symmetry is $\tilde{c}=1-\tilde{j}/4S> 1$, which means that the numerator is always positive on $\tilde{c}\in[-1,1)$. Thus, $f_2(\tilde{j},p_2)$ decreases monotonically on $p_2\in(0,\pi)$ and approaches a positive value as $p_2\to\pi^-$, and hence there is no solution on $I_{\mathrm{R}}$ [Fig.~\ref{N40tan}(b), red and blue curves].
\par ii) $-4S<\tilde{j}<0$.
\par In this case, $f_2(\tilde{j},p_2)$ is singular at $p^*_2=\arccos(-\tilde{j}/4S)$. It is easy to see that $f_2(\tilde{j},p_2)$ is negative on $p_2\in(0,p^*_2)$ and positive on $p_2\in(p^*_2,\pi]$, so there is still no real solutions on $I_{\mathrm{R}}$. Note also that
\begin{eqnarray}
\lim_{p_2\to p^{*\pm}_2}f_2(\tilde{j},p_2)&=&\pm\infty,\nonumber\\
\lim_{p_2\to 0^+}f_2(\tilde{j},p_2)&=&-\infty.
\end{eqnarray}
It can be further shown that $f_2(\tilde{j},p_2)$ is a monotonically decreasing function on $p_2\in(p^*_2,\pi]$. Actually, at $\tilde{c}^*=-\tilde{j}/4S$, the numerator in Eq.~(\ref{partialf2}) is $4S(\tilde{j}+4S)>0$, so $\partial_{\tilde{c}}f_2(\tilde{j},\tilde{c})$ is always positive on $\tilde{c}\in (-1,\tilde{c}^*)$.
\par If $p^*_2\in I_{\mathrm{L}}$, then there is a single solution on $\left(p^*_2,\frac{2\pi}{N-4}\right)$ but no solution on $I_{\mathrm{R}}$. If $p^*_2\in I_{\mathrm{R}}$, then there is a single solution on $\left(\frac{(N-6)\pi}{N-4},p^*_2\right)$ but no solution on $I_{\mathrm{L}}$. If $p^*_2$ lies in any interval other than $I_{\mathrm{L}}$ and $I_{\mathrm{R}}$, then there will be two solutions in this interval. However, in this case no solutions exist in both $I_{\mathrm{L}}$ and $I_{\mathrm{R}}$ [Fig.~\ref{N40tan}(b), green curve]. Therefore, in any case there are $N/4-1$ real solutions of Eq.~(\ref{Teq2}) when $-4S<\tilde{j}<0$.
\par By combining the results in i) and ii), we reach the conclusion that for any $\tilde{j}<0$ Eq.~(\ref{Teq2}) has exactly $N/4-1$ real solutions. As a result, there is a single complex solution $p_2=m\pi+i\tilde{p}_2$ for all $J/J'<0$ and $\Delta>0$. We next show that this complex solution corresponds to the other type of two-magnon bound state with a lower excitation energy. Inserting $p_2=m\pi+i\tilde{p}_2$ into Eq.~(\ref{Teq2}) gives
\begin{eqnarray}\label{Teq2modd}
\frac{\tanh(\frac{N}{4}-1)\tilde{p}_2}{\sinh \tilde{p}_2}&=& \frac{-\tilde{j} +4S(1+\cosh \tilde{p}_2)}{(1+\cosh \tilde{p}_2)(\tilde{j}-4S\cosh \tilde{p}_2)},\nonumber\\
\end{eqnarray}
for odd $m$, and
\begin{eqnarray}\label{Teq2meven}
\frac{\tanh(\frac{N}{4}-1)\tilde{p}_2}{\sinh \tilde{p}_2}&=&\frac{\tilde{j} -4S(1-\cosh \tilde{p}_2)}{(1-\cosh \tilde{p}_2)(4S\cosh \tilde{p}_2+\tilde{j})},\nonumber\\
\end{eqnarray}
for even $m$. It is obvious that Eq.~(\ref{Teq2modd}) has no solution on $\tilde{p}_2\in(0,\infty)$ since the right-hand side is always negative. Thus, $m$ must be an even integer, which leads to the wavenumber-dependent part of the excitation energy $4SJ'\cosh\tilde{p}_2<4SJ'\cos p_2$. For large $N$, the real solutions $\{p_2\}$ tend to be quasi-continuous and form the scattering continuum. We therefore proved that the bound state lies below the scattering continuum. For $S=1/2$, this bound state is the lowest one in the $k=-\pi$ sector since both the continuum and the high-lying NNN-Ex bound state have higher excitation energies. Q.E.D.
\par We now discuss the solution of Eq.~(\ref{Teq2meven}) and the related two-magnon bound state. First note that Eq.~(\ref{Teq2meven}) has no solution near $\tilde{p}_2=0$ since the right-hand side diverges as $\tilde{p}_2\to 0^+$. For large $N$, we thus have $\tanh\left(\frac{N}{4}-1\right)\tilde{p}_2\to 1$ and Eq.~(\ref{Teq2meven}) is reduced to a quadratic equation of $\cosh\tilde{p}_2$,
\begin{eqnarray}
(\cosh\tilde{p}_2-1)[4S(\tilde{j}-2S)(\cosh\tilde{p}_2-1)+\tilde{j}^2]&=&0,
\end{eqnarray}
giving (discarding the unphysical solution $\cosh\tilde{p}_2=1$)
\begin{eqnarray}
\cosh\tilde{p}_2=1+\frac{\tilde{j}^2}{4S(2S-\tilde{j})}.
\end{eqnarray}
Thus, for large $N$ the excitation energy of this bound state is
\begin{eqnarray}\label{Ebound}
E^{(2)}_{\mathrm{NN-Ex}}&=&4SJ\Delta+4SJ'(1+\Delta')+2D(2S-1)+2B\nonumber\\
&&+\frac{ (J\Delta)^2}{  2SJ'-J\Delta },
\end{eqnarray}
where we have restored finite $D$ for $S>1/2$. For $S=1/2$ the above equation becomes
\begin{eqnarray}\label{Eboundq}
E^{(2)}_{\mathrm{NN-Ex}}= 2(J\Delta+B)+2J'(1+\Delta')+\frac{(J\Delta)^2}{J'-J\Delta},
\end{eqnarray}
which is consistent with previous literature~\cite{Chubukov1991,Kuzian2007}.
\par From Eqs.~(\ref{SL2}) and (\ref{vj_p}), we get the corresponding eigenvector,
\begin{eqnarray}
V^{(2)}_{\mathrm{NN-Ex},j}=\cosh[(N/4+1/2-j)\tilde{p}_2],
\end{eqnarray}
with $j=2,\cdots,N/4-1$, and
\begin{eqnarray}
V^{(2)}_{\mathrm{NN-Ex},1}&=&[1-\tilde{j}/(2S)]V^{(2)}_{\mathrm{NN-Ex},1},\nonumber\\
V^{(2)}_{\mathrm{NN-Ex},N/4}&=&\frac{2S(2S-\tilde{j})}{\tilde{j}^2-2S\tilde{j}+4S^2}V^{(2)}_{\mathrm{NN-Ex},N/4-1}.
\end{eqnarray}
Note that $1-\tilde{j}/2S>1$ and $0<2S(2S-\tilde{j})/(\tilde{j}^2-2S\tilde{j}+4S^2)<1$, we have
\begin{eqnarray}
|V^{(2)}_{\mathrm{NN-Ex},1}|>|V^{(2)}_{\mathrm{NN-Ex},2}|>\cdots>|V^{(2)}_{\mathrm{NN-Ex},N/4}|,
\end{eqnarray}
indicating that the state is localized around the left end of the $L_2$ chain (i.e., the site $|\xi_1(-\pi)\rangle$) and corresponds to a two-magnon bound state with the two spin deviations being mainly located on two NN sites in real space. This bound state is the usual nearest-neighbor exchange (NN-Ex) bound state with the two spin derivations mainly located on two nearest-neighboring sites~\cite{PRB2022,PRB1987}.
\par In summary, we proved for $S=1/2$ that in the $k=-\pi$ sector the NNN-Ex two-magnon bound state emerges as the highest excited state when $\Delta'>1$. For any $S\geq 1/2$, the NN-Ex two-magnon bound state always survives below the scattering continuum. From continuous considerations, these properties will persist near the band edge, explaining the presence of the lowest-lying level shown in Fig.~\ref{Jhalfkeck}.
\subsection{$S>1/2$}
\begin{figure}
\includegraphics[width=0.53\textwidth]{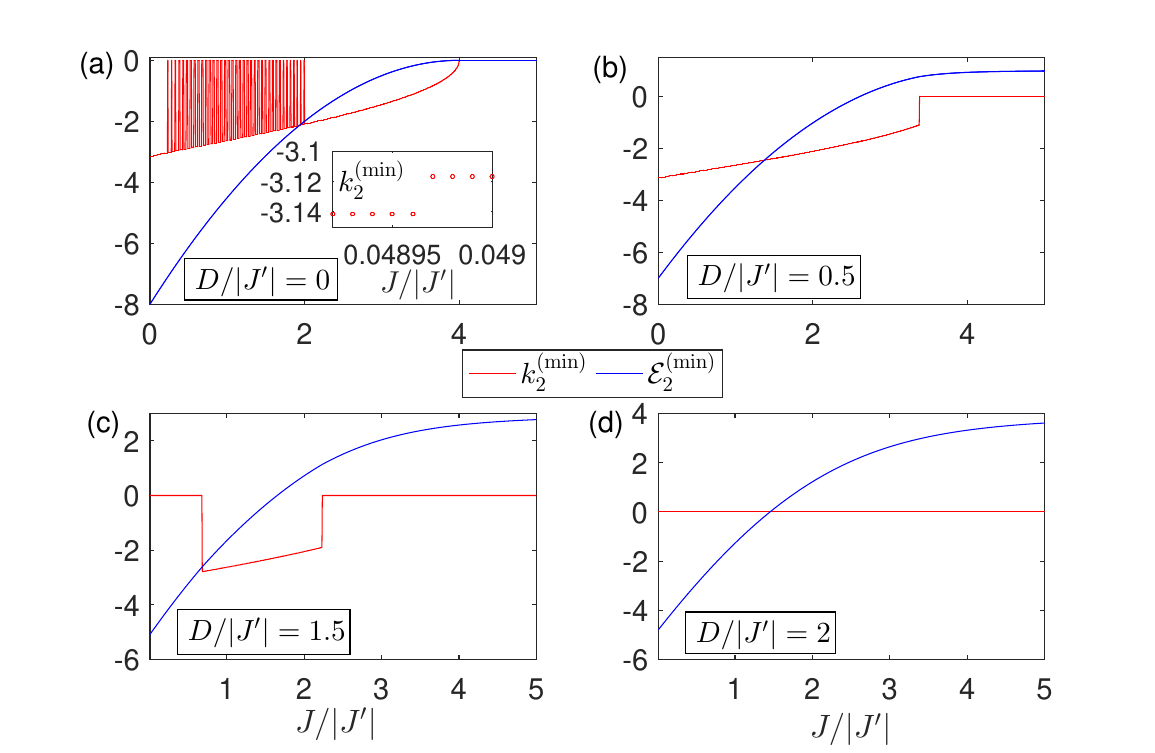}
\caption{Evolution of the lowest two-magnon excitation energy $\mathcal{E}^{(\min)}_2/|J'|=\mathcal{E}_2(k^{(\min)}_2)/|J'|$ (blue) and the corresponding wave number $k^{(\min)}_2$ (red) with increasing $J/|J'|$ for $N=500$ and $S=1$. (a) $D/|J'|=0$, (b) $D/|J'|=0.5$, (c) $D/|J'|=1.5$, (d) $D/|J'|=2$. The inset in (a) shows $k^{(\min)}_2$ around the C-NC transition point $J/|J'| \approx 0.04896$. Other parameters: $\Delta=\Delta'=1$ and $B=0$.}
\label{N500S1}
\end{figure}
\begin{figure*}
\includegraphics[width=1\textwidth]{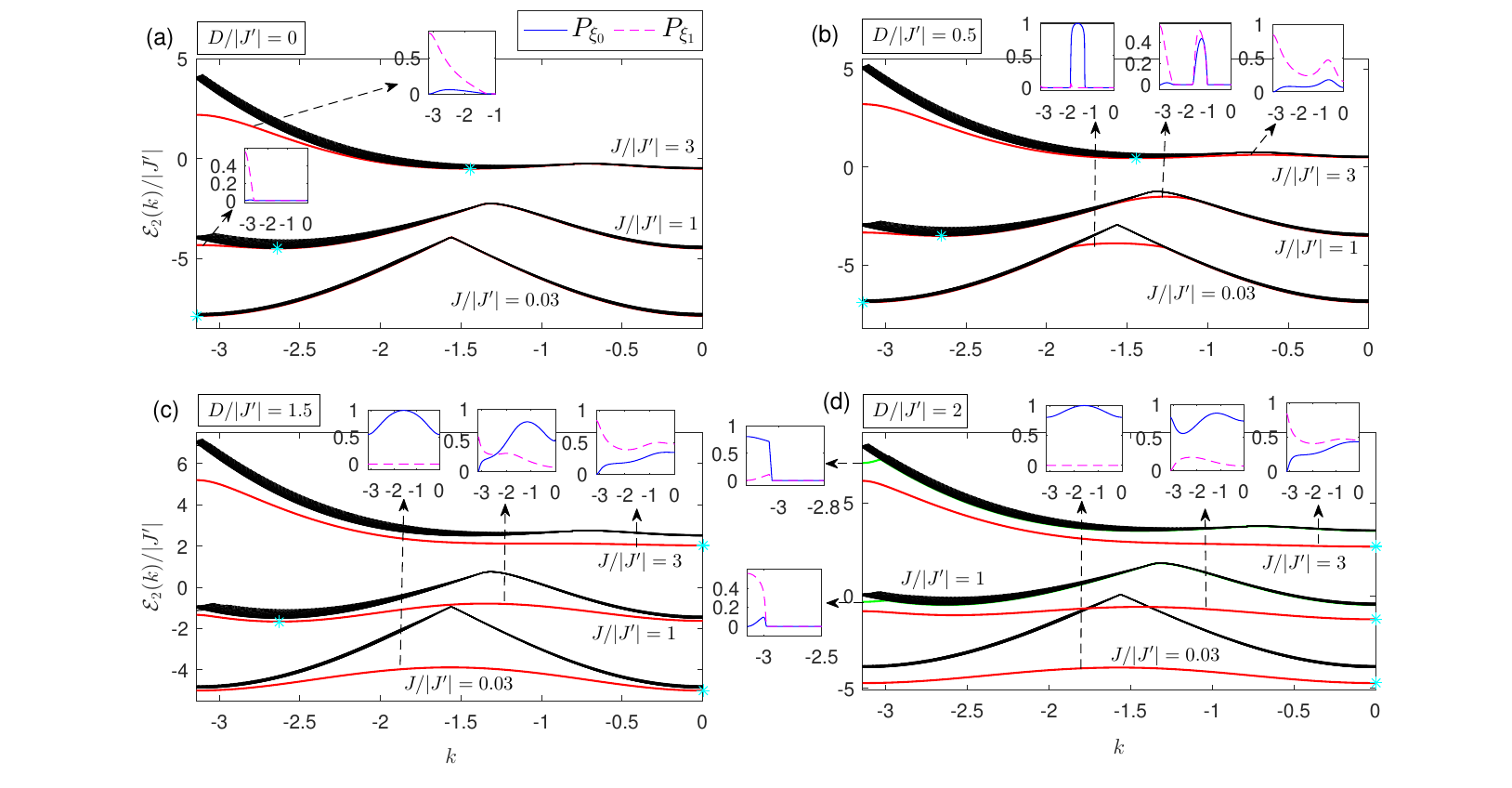}
\caption{The lowest 20 excitation levels $\mathcal{E}_2(k)/|J'|$ for $N=500$ and $S=1$ with varying $D/|J'|$ and $J/|J'|$. (a) $D/|J'|=0$, (b) $D/|J'|=0.5$, (c) $D/|J'|=1.5$, (d) $D/|J'|=2$. In each panel the results for $J/|J'|=0.03$, $1$, and $3$ are shown. The bottom of each lowest level $|\psi_{\mathrm{LL}}(k)\rangle$ is highlighted by a cyan star. The insets show the weight of the two Bloch states $|\xi_0(k)\rangle$ and $|\xi_1(k)\rangle$ in the lowest level, i.e., $P_{\xi_i}(k)=|\langle\xi_i(k)|\psi_{\mathrm{LL}}(k)\rangle|^2$, $i=0,1$. Other parameters: $\Delta=\Delta'=1$ and $B=0$.}
\label{lowspecDJ}
\end{figure*}
\par Let us now turn to study the case of higher spins. We allow for finite values of the SI anisotropy $D$. We first focus on the isotropic case with $\Delta=\Delta'=1$. We plot in Fig.~\ref{N500S1} the evolution of $\mathcal{E}_2^{(\min)}/|J'|$ and $k_2^{(\min)}$ with increasing $J/|J'|$ for $N=500$, $S=1$, $\Delta=\Delta'=1$, and $B=0$.
\par For $D=0$, according to Eq.~(\ref{JnTL}), the minimal two-magnon excitation energy exactly vanishes when $J/|J'|>(J/|J'|)^{(2)}_{\mathrm{th}}(500)\approx 3.999526$. For $0< J/|J'|< (J/|J'|)^{(2)}_{\mathrm{th}}(500)$, the behaviors of $\mathcal{E}_2^{(\min)}/|J'|$ and $k_2^{(\min)}$ are in sharp contrast with those in the case of $S=1/2$ (compared with Fig.~\ref{twomagnonEminKmin}). The C-NC transition point is found to be $\approx 0.04896$ [inset of Fig.~\ref{N500S1}(a)], after which $k_2^{(\min)}$ increases gradually till $J/|J'|=0.229$ where $k_2^{(\min)}$ jumps to $0$. Interestingly, the value of $k_2^{(\min)}$ fluctuates between zero and finite incommensurate values in the middle region $J/|J'|\in(0.229,2.039)$, though the minimal excitation energy $\mathcal{E}_2^{(\min)}/|J'|$ is always smooth. The sudden jump and fluctuation of $k_2^{(\min)}$ are related to the degeneracy of the lowest two excitation levels $\mathcal{E}_2^{(\min)}/|J'|$ and $\mathcal{E}_2^{(\mathrm{2nd}-\min)}/|J'|$ for certain values of $J/|J'|$. The corresponding two local minima are located near the band edge and at $k=0$ [see Fig.~\ref{lowspecDJ}(a)]. As $J/|J'|$ is varying in the middle region, one of the two local minima alternately becomes the global minimum, causing the observed fluctuation of $k_2^{(\min)}$. However, numerical tests show that the difference between the two levels $[\mathcal{E}_2^{(\mathrm{2nd}-\min)}-\mathcal{E}_2^{(\min)}]/(N|J'|)$ tends to be vanishingly small as $N\to\infty$. We thus believe that this phenomenon is a finite-size effect and will disappear in the thermodynamic limit.
\par To see the nature of the lowest excitation, we plot in Fig.~\ref{lowspecDJ}(a) the lowest 20 excitation levels for $J/|J'|=0.03$, $1$, and $3$. The bottom of the lowest-lying level $|\psi_{\mathrm{LL}}(k)\rangle$ is indicated by a cyan star. These bottom states all correspond to two-magnon scattering states for $D/|J'|=0$, as can be seen from the evolution of the weights of the Bloch states $|\xi_i(k)\rangle$ ($i=0,1$) with increasing $k$, $P_{\xi_i}(k)=|\langle\xi_i(k)|\psi_{\mathrm{LL}}(k)\rangle|^2$. Note that for larger $J/|J'|$ the NN-Ex bound states will emerge as a lower separated level near the band edge $k=-\pi$ [inset of Fig.~\ref{lowspecDJ}(a)].
\par Figure~\ref{N500S1}(b) shows $\mathcal{E}_2^{(\min)}/|J'|$ and $k_2^{(\min)}$ for $D/|J'|=0.5$. It can be seen that as $J/|J'|$ increases, $k_2^{(\min)}$ no longer shows fluctuations but increases gradually from $-\pi$ to $-1.1058$ at $J/|J'|=3.375$, where $k_2^{(\min)}$ suddenly jumps to $k_2^{(\min)}=0$. The bottom states are still scattering states for small $J/|J'|$ [inset of Fig.~\ref{lowspecDJ}(b)]. However, besides the NN-Ex bound state near the edges of the band for larger $J/|J'|$, the so-called single-ion (SI) bound states~\cite{PRB2022,PRB1987} with the two spin deviations located on a single site also appear in the middle of the band for smaller $J/|J'|$.
\par For $D/|J'|=1.5$, $k_2^{(\min)}$ never reaches $-\pi$ and is nonzero in the interval $J/|J'|\in(0.685,2.23)$ [Fig.~\ref{N500S1}(c)]. The bottom mode for $J/|J'|=0.03$ is $k_2^{(\min)}=0$ and the corresponding state is an SI bound state. However, for larger $J/|J'|$ the bottom state is a mixture of the NN-Ex and SI bound states [inset of Fig.~\ref{lowspecDJ}(c)].
\par As $D/|J'|$ increases to $2$, we observe that $k_2^{(\min)}$ is always zero [Fig.~\ref{N500S1}(d)] and the corresponding bottom states are SI bound states for not too large $J/J'$. For $J/J'=3$, the lowest state evolves from the NN-Ex bound state to the mixture of the two as $k$ increases. In addition, we observe that for $J/|J'|=1$ ($J/|J'|=3$) a second separated level near $k=-\pi$ emerges as an NN-Ex (an SI) bound state [inset of Fig.~\ref{lowspecDJ}(d)].
\par To see more clearly how the three types of bound states emerge at the left edge of the band, it is instructive to study the special mode $k=-\pi$ for which the problem can also be solved via the plane-wave ansatz. Since the $L_2$ chain has been solved in Sec.~\ref{SecIVA} for arbitrary $S\geq 1/2$, here we focus on the solution of the $L_1$ chain.  For $S>1/2$, let the eigenvectors be
\begin{eqnarray}
V^{(1)}&=&(V^{(1)}_1,\cdots,V^{(1)}_{N/4+1})^T,\nonumber\\
V^{(1)}_j&=&X_1 e^{ip_1 j}+Y_1 e^{-ip_1 j},~j=2,\cdots,\frac{N}{4}.
\end{eqnarray}
The eigenenergies are given by
\begin{eqnarray}
E^{(1)}(p_1)&=&4S(J\Delta+J'\Delta')+2D(2S-1)+2B\nonumber\\
&&+4SJ'\cos p_1.
\end{eqnarray}
According to Eq.~(\ref{Treq}), the wave number $p_1$ satisfies the following equation
\begin{eqnarray}\label{Teq1S1a}
\tan\frac{Np_1}{4}&=&g_1(\Delta',d,p_1),\nonumber\\
g_1(\Delta',d,p_1) &=&\frac{w^{(+)}(\cos p_1)}{(A-\cos p_1)\sin p_1},
\end{eqnarray}
where $d\equiv D/|J'|>0$, $A\equiv (2S-1)/\Delta'+d/(2S)>0$, and
\begin{eqnarray}
w^{(\pm)}(x)\equiv x^2\pm(2S/\Delta'-A)x-d/\Delta'.
\end{eqnarray}
The SI and NNN-Ex bound states, if exist, will show up in $L_1$ and depend on both $d$ and $\Delta'$. We now discuss the solutions of Eq.~(\ref{Teq1S1a}).
\begin{figure}
\includegraphics[width=0.52\textwidth]{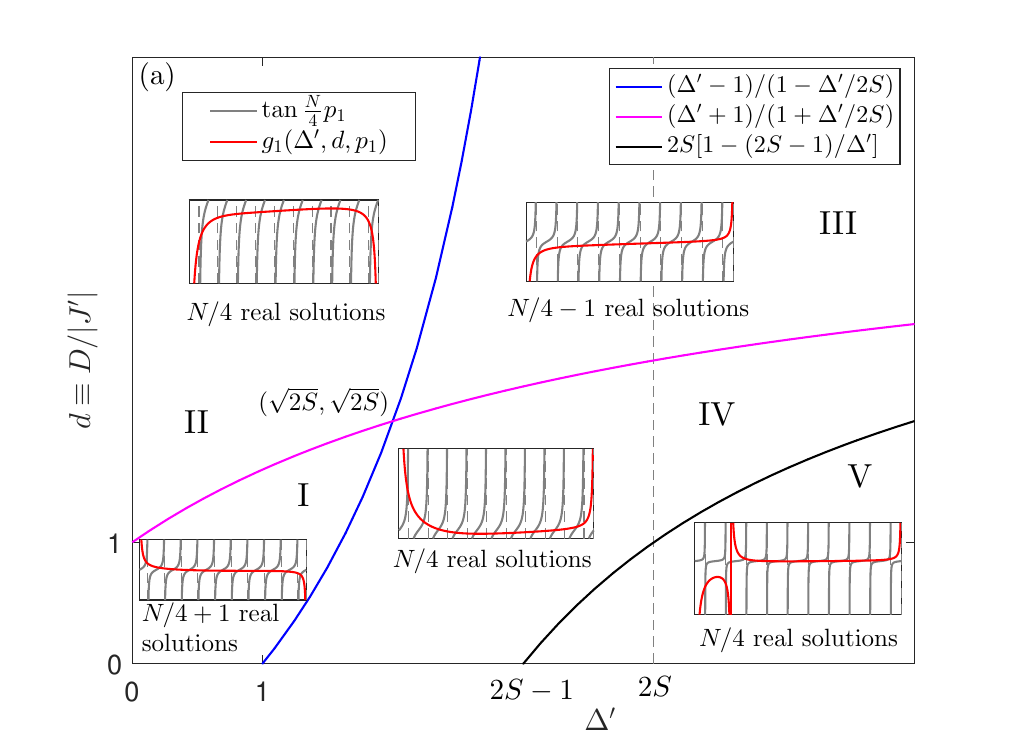}
\includegraphics[width=0.52\textwidth]{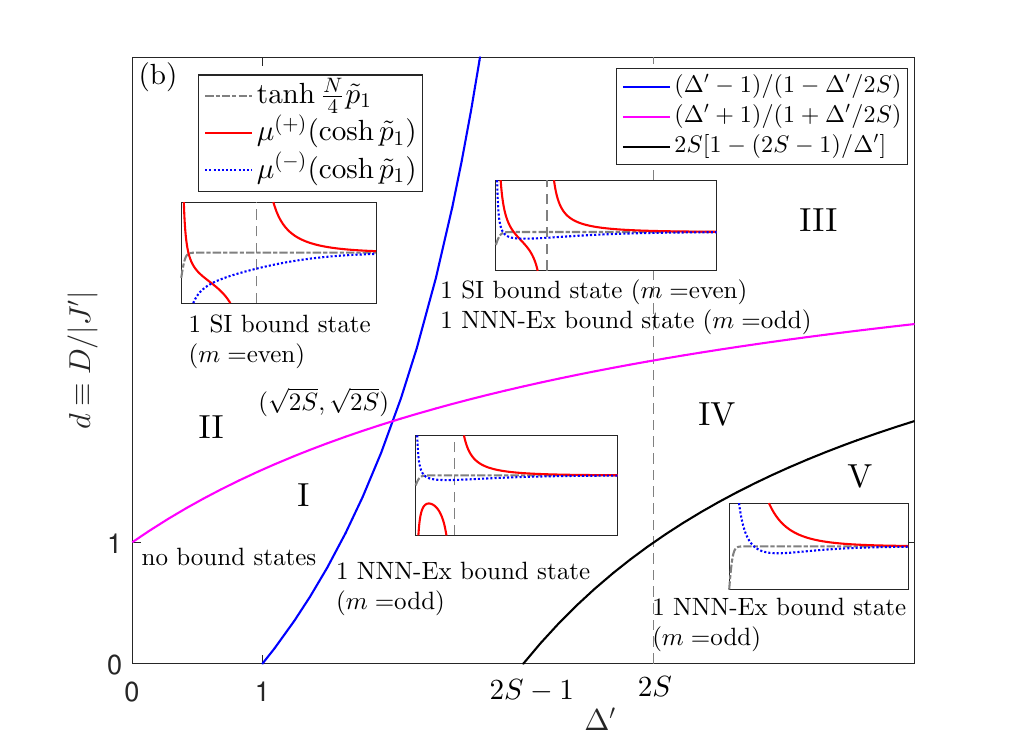}
\caption{The first quadrant of the $\Delta'-d$ (where $d\equiv D/|J'|$) plane is divided into five regions I, II, III, IV, and V by the three functions $2S[1-(2S-1)/\Delta']$, $(\Delta'+1)/(1+\Delta'/2S)$, and $(\Delta'-1)/(1-\Delta'/2S)$. The solutions of Eq.~(\ref{Teq1S1a}) have different structures in district regions. (a) Intersections of the graphs of $\tan\frac{N}{4}p_1$ and $g_1(\Delta',d,p_1)$ give the real solutions of Eq.~(\ref{Teq1S1a}) in each region. (b)  Intersections of the graphs of $\tanh\frac{N}{4}\tilde{p}_1$ and $\mu^{(\pm)}(\cosh\tilde{p}_1)$ give the complex solutions of Eq.~(\ref{Teq1S1a}) [or the real solutions of Eqs.~(\ref{Teq1S1aTanhodd}) and (\ref{Teq1S1aTanheven})] in each region. Accordingly, different types of two-magnon bound states emerge in different regions.}
\label{S2Ddel}
\end{figure}
\par i) $A>1$, or
\begin{eqnarray}
d>2S[1-(2S-1)/\Delta'].
\end{eqnarray}
In this case, $g_1(\Delta',d,p_1)$ has no singularity as a function of $p_1$. The behavior of $g_1(\Delta',d,p_1)$ near $p_1=0$ or $\pi$ depends on the sign of $w^{(+)}(1)$ or $w^{(+)}(-1)$.
\par i-a) $w^{(+)}(1)>0$ and $w^{(+)}(-1)>0$, or
\begin{eqnarray}
d<\frac{\Delta'+1}{1+\Delta'/(2S)},
\end{eqnarray}
and
\begin{eqnarray}
d<\frac{\Delta'-1}{1-\Delta'/(2S)}~(\mathrm{with}~\Delta'<2S)~\mathrm{or}~\Delta'>2S
\end{eqnarray}
The above two inequalities defines the region IV in the first quadrant of the $\Delta'-d$ plane, as shown in Fig.~\ref{S2Ddel}(a). In this region we have
\begin{eqnarray}
\lim_{p_1\to 0^+}g_1(\Delta',d,p_1)=+\infty,~\lim_{p_1\to \pi^-}g_1(\Delta',d,p_1)=+\infty,\nonumber
\end{eqnarray}
see the insets of Fig.~\ref{S2Ddel}(a) where we plotted the graphs of the two functions $\tan\frac{N}{4}p_1$ and $g_1(\Delta',d,p_1)$ in each region. In turn, there are $N/4$ real solutions and a single complex solution in region IV.
\par i-b) $w^{(+)}(1)>0$ and $w^{(+)}(-1)<0$.
\par Similar analysis shows that these conditions define the region I in Fig.~\ref{S2Ddel}(a), where we have
\begin{eqnarray}
\lim_{p_1\to 0^+}g_1(\Delta',d,p_1)=+\infty,~\lim_{p_1\to \pi^-}g_1(\Delta',d,p_1)=-\infty.\nonumber
\end{eqnarray}
There are thus $N/4+1$ real solutions and no complex solution in region I.
\par i-c) $w^{(+)}(1)<0$ and $w^{(+)}(-1)>0$.
\par These conditions define the region III in Fig.~\ref{S2Ddel}(a) with
\begin{eqnarray}
\lim_{p_1\to 0^+}g_1(\Delta',d,p_1)=-\infty,~\lim_{p_1\to \pi^-}g_1(\Delta',d,p_1)=+\infty.\nonumber
\end{eqnarray}
There are thus $N/4-1$ real solutions and two complex solutions in region III.
\par i-d) $w^{(+)}(1)<0$ and $w^{(+)}(-1)<0$.
\par These conditions define the region II in Fig.~\ref{S2Ddel}(a) with
\begin{eqnarray}
\lim_{p_1\to 0^+}g_1(\Delta',d,p_1)=-\infty,~\lim_{p_1\to \pi^-}g_1(\Delta',d,p_1)=-\infty.\nonumber
\end{eqnarray}
There are thus $N/4$ real solutions and a single complex solutions in region II.
\par ii) $0<A<1$.
\par This inequality defines the region V in Fig.~\ref{S2Ddel}(a). The function $g_1(\Delta',d,p_1)$ is singular at $p^*_1=\arccos A$. It is obvious that $w^{(+)}(1)>0$ and $w^{(+)}(-1)>0$ in region V. However, we also have $\lim_{p_1\to0^+}(A-\cos p_1)<0$ and $\lim_{p_1\to\pi^-}(A-\cos p_1)>0$, giving
\begin{eqnarray}
\lim_{p_1\to 0^+}g_1(\Delta',d,p_1)=-\infty,~\lim_{p_1\to \pi^-}g_1(\Delta',d,p_1)=+\infty.\nonumber
\end{eqnarray}
Note now that $w^{(+)}(A)=4S^2(2S-1)/\Delta'>0$, we have
\begin{eqnarray}
\lim_{p_1\to p^{*\pm}_1}g_1(\Delta',d,p_1)=\pm\infty.\nonumber
\end{eqnarray}
Therefore, there are always $N/4$ real solutions and a single complex solution in region V.
\par Let us now pursue complex solutions of Eq.~(\ref{Teq1S1a}) in regions II, III, IV, and V. We again write the complex solution as $p_1=m\pi+i\tilde{p}_1$ with $m$ an integer and $\tilde{p}_1>0$ real. By defining the functions
\begin{eqnarray}\label{mu}
\mu^{(\pm)}(x)\equiv\mp\frac{w^{(\pm)}(x)}{(A\mp x)\sqrt{x^2-1}}~~(x>1),
\end{eqnarray}
equation (\ref{Teq1S1a}) becomes
\begin{eqnarray}\label{Teq1S1aTanhodd}
\tanh\frac{N\tilde{p}_1}{4}=\mu^{(-)}(\cosh\tilde{p}_1) %\frac{w^{(-)}(\cosh\tilde{p}_1)}{(A+\cosh \tilde{p}_1)\sinh \tilde{p}_1},
\end{eqnarray}
for odd $m$ and
\begin{eqnarray}\label{Teq1S1aTanheven}
\tanh\frac{N\tilde{p}_1}{4}=\mu^{(+)}(\cosh\tilde{p}_1)%-\frac{w^{(+)}(\cosh\tilde{p}_1)}{(A-\cosh \tilde{p}_1)\sinh \tilde{p}_1}
\end{eqnarray}
for even $m$. There is no solution near $\tilde{p}_1=0$, so for large $N$ we have $\tanh N\tilde{p}_1/4\approx 1$ and the above two equations are approximated to cubic equations of $\cosh\tilde{p}_1$.
\par The functions $\mu^{(\pm)}(x)$ have several useful properties. It is obvious that
\begin{eqnarray}\label{muinf}
\lim_{x\to+\infty}\mu^{(\pm)}(x)=1^\pm.
\end{eqnarray}
From the signs of $w^{(\pm)}(1)$ in each region, we get
\begin{eqnarray}\label{mup1}
\lim_{x\to 1^+}\mu^{(+)}(x)&=& \begin{cases}
+\infty, & (\Delta',d)\in\mathrm{II}, \mathrm{III}, \mathrm{V}\\
-\infty, & (\Delta',d)\in\mathrm{IV},
  \end{cases}
\end{eqnarray}
and
\begin{eqnarray}\label{mum1}
\lim_{x\to 1^+}\mu^{(-)}(x)&=& \begin{cases}
+\infty, & (\Delta',d)\in\mathrm{III}, \mathrm{IV}, \mathrm{V}\\
-\infty, & (\Delta',d)\in\mathrm{II}.
  \end{cases}
\end{eqnarray}
In regions $\mathrm{II}$, $\mathrm{III}$, and $\mathrm{IV}$ where $A>1$, the function $\mu^{(+)}(x)$ is singular at $\tilde{p}^*_1$ with $\cosh \tilde{p}^*_1=A$. From the relation $w^{(+)}(\cosh \tilde{p}^*_1)=4S^2(2S-1)/\Delta'>0$, we have
\begin{eqnarray}\label{muppstar}
\lim_{x\to A^{\pm}}\mu^{(+)}(x)&=& \pm\infty,~(\Delta',d)\in\mathrm{II}, \mathrm{III}, \mathrm{IV}.
\end{eqnarray}
The behaviors of $\mu^{(\pm)}(\cosh\tilde{p}_1)$ described by Eqs.~(\ref{muinf})-(\ref{muppstar}) are illustrated in the insets of Fig.~\ref{S2Ddel}(b). The solutions of Eqs.~(\ref{Teq1S1aTanhodd}) and (\ref{Teq1S1aTanheven}) can be determined by investigating the graphs of the related functions shown in Fig.~\ref{S2Ddel}(b).
\par In region II, there exists a single intersection of $\tanh N\tilde{p}_1/4$ and $\mu^{(+)}(\cosh\tilde{p}_1)$ (with even $m$) on $\tilde{p}_1\in (0,A)$. %, and there is no intersection of $\tanh N\tilde{p}_1/4$ and $\mu^{(+)}(\cosh\tilde{p}_1)$.
We thus get a single two-magnon bound state with excitation energy
\begin{eqnarray}\label{EboundSI}
E^{(1)}_{\mathrm{SI}}&=&4S(J\Delta+J'\Delta')+2D(2S-1)+2B\nonumber\\
&&+4SJ'\cosh \tilde{p}_1,
\end{eqnarray}
which lies below the continuum since $\cosh \tilde{p}_1>\cos p_1$.
\par From Eq.~(\ref{SL1}) and (\ref{vj_p}), we get the eigenvector
\begin{eqnarray}
V^{(1)}_{\mathrm{SI},j}=\cosh[(N/4+1-j)\tilde{p}_1],
\end{eqnarray}
with $j=2,\cdots,N/4$, and
\begin{eqnarray}
V^{(1)}_{\mathrm{SI},1}&=&\frac{\sqrt{S(2S-1)}}{2S\cosh \tilde{p}_1-d}V^{(2)}_{\mathrm{SI},2},\nonumber\\
V^{(1)}_{\mathrm{SI},N/4+1}&=&\frac{1}{\sqrt{2}\cos\tilde{p}_1}V^{(2)}_{\mathrm{SI},N/4}.
\end{eqnarray}
Although we have $|V^{(1)}_{\mathrm{SI},N/4}|<|V^{(2)}_{\mathrm{SI},N/4+1}|$, it is not straightforward to see $|\sqrt{S(2S-1)}/(2S\cosh \tilde{p}_1-d)|>1$. However, based on both physical considerations and numerical tests, we find that this is the case, indicating that the state is indeed an SI two-magnon bound state.
\par In region III, the function $\tanh N\tilde{p}_1/4$ intersects with both $\mu^{(+)}(\cosh\tilde{p}_1)$ (with even $m$) and $\mu^{(-)}(\cosh\tilde{p}_1)$ (with odd $m$) on $\tilde{p}_1\in (0,A)$. The former still corresponds to an SI bound state, while the latter leads to an NNN-Ex two-magnon bound state with excitation energy
\begin{eqnarray}
E^{(1)}_{\mathrm{NNN-Ex}}&=&4S(J\Delta+J'\Delta')+2D(2S-1)+2B\nonumber\\
&&-4SJ'\cosh \tilde{p}_1,
\end{eqnarray}
which lies above the continuum. The corresponding eigenvector reads (for $j=2,\cdots,N/4$)
\begin{eqnarray}
V^{(1)}_{\mathrm{NNN-Ex},j}=(-1)^j\cosh[(N/4+1-j)\tilde{p}_1],
\end{eqnarray}
and
\begin{eqnarray}
V^{(1)}_{\mathrm{NNN-Ex},1}&=&-\frac{\sqrt{S(2S-1)}}{2S\cosh \tilde{p}_1+d}V^{(2)}_{\mathrm{NNN-Ex},2},\nonumber\\
V^{(1)}_{\mathrm{NNN-Ex},N/4+1}&=&-\frac{1}{\sqrt{2}\cos\tilde{p}_1}V^{(2)}_{\mathrm{NNN-Ex},N/4}.
\end{eqnarray}
It can be numerically checked that $-1<-\sqrt{S(2S-1)}/(2S\cosh \tilde{p}_1+d)<0$, confirming that this bound is actually an NNN-Ex bound state.
\par Similar analysis shows that regions IV and V both support NNN-Ex two-magnon bound states. The phase diagram in the $\Delta'-d$ plane is summarizes in Fig.~\ref{S2Ddel}(b). Recall that we have proved in Sec.~\ref{SecIVA} that the NN-Ex bound state shows up as a lower-lying level in all the regions of the phase diagram, the lowest excited state could be determined by comparing the $E^{(2)}_{\mathrm{NN-Ex}}$ given by Eq.~(\ref{Ebound}) and $E^{(1)}_{\mathrm{SI}}$ given by Eq.~(\ref{EboundSI}).
The above results for $k=-\pi$ are believed to faithfully reflect the nature of two-magnon excitations near the edge of the Brillouin zone.
\section{$n$-magnon excitations for $S=1/2$}\label{SecV}
\par In this section we proceed to $n$-mangon excitations with $n\geq 3$. The exact three-magnon Bloch states and the associated Bloch Hamiltonians for a finite-size spin-$S$ XXZ chain (with $J'=\Delta'=0$) have been constructed in Ref.~\cite{PRB2022}. The derivation of the Bloch Hamiltonians for the NNN interaction is straightforward though cumbersome and will be presented in a future work.
\par In this section, we focus on the case of $S=1/2$ for which the nearest-neighboring XX chain $H_{\mathrm{XX}}=\sum^N_{j=1}(S^x_jS^x_{j+1}+S^y_jS^y_{j+1})$ is analytically soluble by converting the Pauli operators into spinless fermions. The matrix elements of each term in $H$ can be expressed in terms of the so-called spin-operator matrix elements in the diagonal basis of $H_{\mathrm{XX}}$~\cite{PRB2018}. Explicitly, let $|\vec{\eta}_n\rangle$ be an eigenstate of $H_{\mathrm{XX}}$ having $n$ fermions upon the vacuum state $|\downarrow\cdots\downarrow\rangle$, where $\vec{\eta}_n=(\eta_1,\cdots,\eta_n)$ is a tuple with $1\leq\eta_1<\cdots<\eta_n\leq N$, then
\begin{widetext}
\begin{eqnarray}\label{Hxyr}
&&\langle\vec{\chi}_n|\sum_j(S^x_jS^x_{j+r}+S^y_jS^y_{j+r})|\vec{\chi}'_n\rangle\nonumber\\
&=&\left(\frac{2}{N}\right)^{2(n-1)}\delta(\Delta_{\vec{\chi}_n,\vec{\chi}'_n},0)\sum_{\vec{\xi}_{n-1}} A^*_{\vec{\chi}_n}e^{ir\sum_j Q^{(\sigma_n)}_{\chi_j}} C^*_{\vec{\chi}_n,\vec{\xi}_{n-1}} |A_{\vec{\xi}_{n-1}}|^2e^{-ir\sum_j Q^{(\sigma_{n-1})}_{\xi_{j}}} A_{\vec{\chi}'_n}C_{\vec{\chi}'_n,\vec{\xi}_{n-1}}+\mathrm{c.c.},
\end{eqnarray}
\begin{eqnarray}\label{Hzr}
&&\langle\vec{\chi}_n|\sum_jS^z_jS^z_{j+r}|\vec{\chi}'_n\rangle=\left(\frac{N}{4}-n\right)\delta_{\vec{\chi}_n,\vec{\chi}'_n} +\frac{\delta(\Delta_{\vec{\chi}_n,\vec{\chi}'_n},0)}{N}\left(\frac{2}{N}\right)^{4(n-1)}\sum_{\vec{\eta}_n,\vec{\xi}_{n-1},\vec{\xi}'_{n-1}}\nonumber\\
&&\left(A^*_{\vec{\chi}_n}e^{ir\sum_j Q^{(\sigma_n)}_{\chi_j}}C^*_{\vec{\chi}_n,\vec{\xi}_{n-1}}|A_{\vec{\xi}_{n-1}}|^2\right)\left(C_{\vec{\eta}_n,\vec{\xi}_{n-1}}e^{-ir\sum_j Q^{(\sigma_n)}_{\eta_j}}|A_{\vec{\eta}_n}|^2C^*_{\vec{\eta}_n,\vec{\xi}'_{n-1}}\right)\left(|A_{\vec{\xi}'_{n-1}}|^2C_{\vec{\chi}'_n,\vec{\xi}'_{n-1}}A_{\vec{\chi}'_n }\right).
\end{eqnarray}
\end{widetext}
In the above equations, $\delta(x,y)=1$ if $x=y~(\mathrm{mod}~2\pi)$, $\Delta_{\vec{\chi}_n,\vec{\chi}'_n}= \sum_j[  Q^{(\sigma_n)}_{\chi_j}-  Q^{(\sigma_n)}_{\chi'_j}]$, where $Q^{(\sigma_n)}_{\chi_j}=-\pi+2[\chi_j+(\sigma_n-3)/2]\pi/N$ with $\sigma_n=1$ (even $n$) or $\sigma_n=-1$ (odd $n$). The explicit expressions for the $A$'s and $C$'s read
\begin{eqnarray} \label{AB}
A_{\vec{\chi}_n}&=&\prod_{j>j'}\left(e^{iQ^{(\sigma_n)}_{\chi_j}}-e^{iQ^{(\sigma_n)}_{\chi_{j'}}}\right),\nonumber\\
C_{\vec{\chi}_n,\vec{\xi}_{n-1}}&=&\left(\frac{i}{2}\right)^{(n-1)n}\prod_{ij}\csc \frac{Q^{(\sigma_n)}_{\chi_j}-Q^{(\sigma_{n-1})}_{\xi_i}}{2}\nonumber\\
&&\times e^{\frac{i}{2}\left[(n-1)\sum_jQ^{(\sigma_n)}_{\chi_j}-n\sum_iQ^{(\sigma_{n-1})}_{\xi_i}\right]}.
\end{eqnarray}
\begin{figure}
\includegraphics[width=0.51\textwidth]{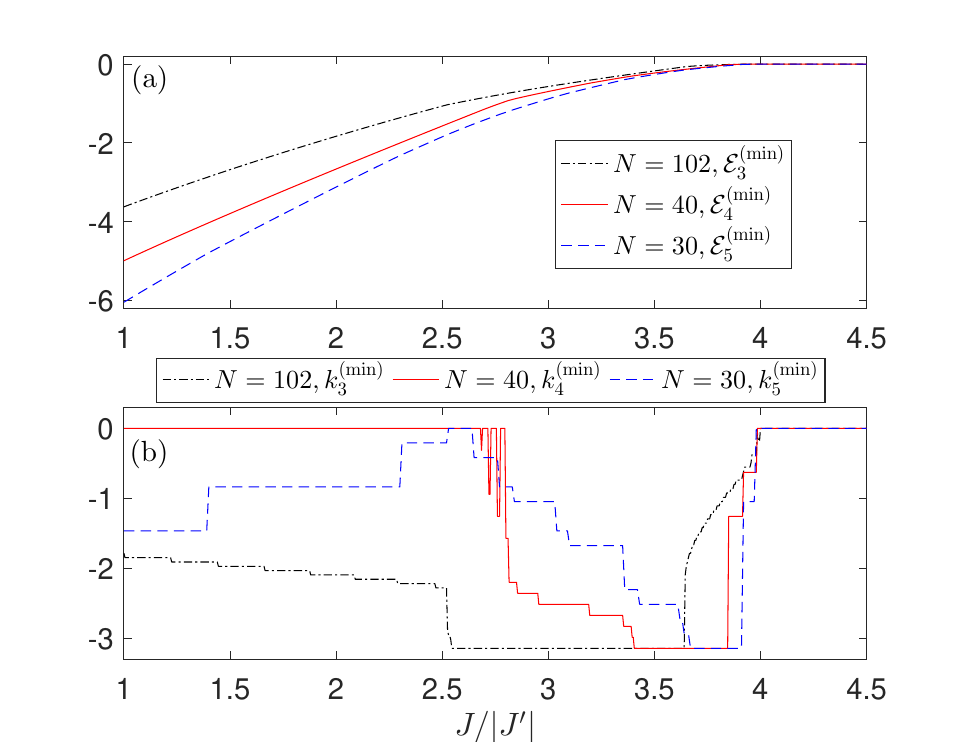}
\caption{(a) The zero-field lowest $n$-magnon excitation energy $\mathcal{E}^{(\min)}_n$ for $S=1/2$. Results for $n=3$ ($N=102$, black dot-dashed), $n=4$ ($N=40$, red solid), and $n=5$ ($N=30$, blue dashed) are shown. (b) The corresponding wave number $k^{(\min)}_n$ at which $\mathcal{E}^{(\min)}_n$ is reached. Parameters: $\Delta=\Delta'=1$ and $B=0$.}
\label{TFFEK}
\end{figure}
\par In practice, the evaluation of the $C$-functions given by Eq.~(\ref{AB}) is the most time-consuming step in the numerics. Due to memory limitations, we choose to numerically calculate the three-, four-, and five-magnon excitation spectra up to $N=102$, $N=40$, and $N=30$, with the dimensions of the Hilbert space being $\binom{102}{3}=171,700$,~$\binom{40}{4}=91,390$, and $\binom{30}{5}=142,506$, respectively. However, notice the translational invariance of the system reflected in the $\delta$-functions, the whole Hilbert space is split into smaller blocks with fixed $k=\sum_j Q^{(\sigma_n)}_{\chi_j}~(\mathrm{mod}~2\pi)$, which can be handled on a personal computer. Below we focus on the isotropic case with $\Delta=\Delta'=1$.
\begin{figure}
\includegraphics[width=0.51\textwidth]{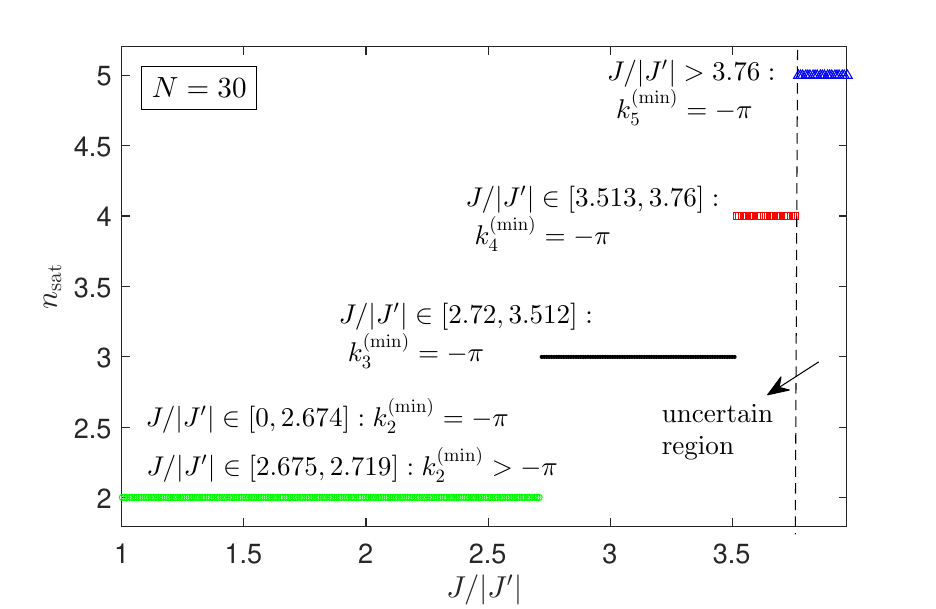}
\caption{The number of magnons $n_{\mathrm{sat}}$ in the lowest excited state when the magnetic field is tuned to the saturated value $B_{\mathrm{sat}}$. For $N=30$, the $n_{\mathrm{sat}}=2\to n_{\mathrm{sat}}=3$, $n_{\mathrm{sat}}=3\to n_{\mathrm{sat}}=4$, and $n_{\mathrm{sat}}=4\to n_{\mathrm{sat}}=5$ transition points with varying $J/|J'|$ are determined to be $J/|J'|=2.719,3.513$, and $3.76$, respectively. The transition $n_{\mathrm{sat}}=5\to n_{\mathrm{sat}}=6$ is expected to take place in the uncertain region $3.76<J/|J'|$ (due to the limitation of the numerics). Parameters: $\Delta=\Delta'=1$.}
\label{Bsat}
\end{figure}
\par Figure~\ref{TFFEK}(a) shows the calculated lowest excitation energies in the three magnetization sectors when the magnetic field $B$ is absent. As expected, for fixed $J/|J'|<4$, we have $\mathcal{E}^{(\min)}_5<\mathcal{E}^{(\min)}_4<\mathcal{E}^{(\min)}_3<\mathcal{E}^{(\min)}_2<\mathcal{E}^{(\min)}_1<0$. The corresponding wave number $k^{(\min)}_n$ is plotted in Fig.~\ref{TFFEK}(b). A detailed numerical analysis reveals that $k^{(\min)}_n=-\pi$ is achieved for $2.544\leq J/|J'|\leq 3.644$ ($n=3$, $N=102$), $3.404\leq J/|J'|\leq 3.849$ ($n=4$, $N=40$), and $3.666\leq J/|J'|\leq 3.918$ ($n=5$, $N=30$).
\par We now consider the case of finite magnetic fields. We define the saturation field $B_{\mathrm{sat}}$ as the magnetic field that makes the lowest excited state gapless~\cite{Furusaki2007}. Suppose this lowest state lies in the $n_{\mathrm{sat}}$-magnon sector, we focus on excitations up to $n=5$ magnons in a chain with $N=30$ sites. Figure~\ref{Bsat} shows $n_{\mathrm{sat}}$ as a function of $J/|J'|$. We find that (for $N=30$) the $n_{\mathrm{sat}}=2\to n_{\mathrm{sat}}=3$, $n_{\mathrm{sat}}=3\to n_{\mathrm{sat}}=4$, and $n_{\mathrm{sat}}=4\to n_{\mathrm{sat}}=5$ transitions occur at $J/|J'|=2.719$, $3.513$, and $3.76$, respectively. These numerically exact results are very close to those obtained in a restricted Hilbert space~\cite{Furusaki2008}. Note that we were not able to determine the $n_{\mathrm{sat}}=5\to n_{\mathrm{sat}}=6$ transition point since $n=6$ is beyond our numerics.
\section{Conclusions and discussions}\label{SecVI}
\par The spin-1/2 $J-J'$ chain with ferromagnetic nearest-neighbor and antiferromagnetic next-nearest-neigbhor couplings has attracted much attention in previous works due to its relevance to real magnetic materials. However, its higher-spin counterpart with the single-ion anisotropy included is less studied. Motivated by recent experimental advances in simulations of higher-spin magnetic models, we study theoretically exact few-magnon excitations in a finite-size spin-$S$ $J-J'$ chain with single-ion anisotropy.
\par As a related problem, we first study the emergence of zero-excitation-energy states in the absence of the single-ion anisotropy and identify the corresponding condition to achieve them. In the isotropic case, we determine the threshold of $J/|J'|$ above which ferromagnetic ground states survive. This threshold is found to be exactly $4$ for $S=1/2$ but show size-dependence for $S>1/2$, which are numerically obtained through exact diagonalization on small systems.
\par We then thoroughly investigate the two-magnon excitations by using a set of exact two-magnon Bloch states proposed for a spin-$S$ XXZ ring~\cite{PRB2022}. We recover prior results for the case of $S=1/2$~\cite{Chubukov1991,Kuzian2007,Furusaki2007,Furusaki2008}. For higher spins, owing to the interplay of the single-ion anisotropy and the NNN exchange coupling, the evolution of the lowest excitation energy and the corresponding wave number with varying $J/|J'|$ exhibit different behaviors from that for $S=1/2$. In particular, we solve the eigenvalue problem of the commensurate mode $k=-\pi$ using a plane-wave ansatz, from which we identify the parameters regions that support the three different types of two-magnon bound states near the band edge. We prove that there always exists lower-energy nearest-neighbor exchange two-magnon bound states near $k=-\pi$.
\par We finally calculate the $n$-magnon spectra for $S=1/2$ using a spin-operator matrix element method. Under the saturation field, the number of magnons in the lowest state takes transitions as $J/|J'|$ is varied. Our numerically exact results for a chain of $N=30$ sites are consistent with those obtained in a restricted Hilbert space~\cite{Furusaki2008}.
\par Considering the possible experimental realization of the present model in cold-atom systems, it is intriguing to study multimagnon quantum walks and related nonequilibrium dynamics in future works.\\
\\
\noindent{\bf Acknowledgements:}
N.W. thanks H. Katsura for useful discussions and suggesting the plane-wave ansatz solution of the effective nearest-neighboring open chain. This work was supported by the National Key Research and Development Program of China under Grant No. 2021YFA1400803.

\end{document}